\documentclass[twocolumn,aps,superscriptaddress,showpacs]{revtex4}
\usepackage{amssymb}
\usepackage{amsmath}
\usepackage{graphicx}
\usepackage[normalem]{ulem}
\usepackage[dvips]{color}

\renewcommand\sout{\bgroup \color{red} \ULdepth=-.5ex \ULset}

\begin{document}

\title{Effects of the $U$-boson on the inner edge of neutron star crusts}
\author{Hao Zheng}
\affiliation{Department of Physics, Shanghai Jiao Tong University, Shanghai 200240, China}
\author{Lie-Wen Chen\footnote{%
Corresponding author (email: lwchen$@$sjtu.edu.cn)}}
\affiliation{Department of Physics, Shanghai Jiao Tong University, Shanghai 200240, China}
\affiliation{Center of Theoretical Nuclear Physics, National Laboratory of Heavy Ion
Accelerator, Lanzhou 730000, China}
\date{\today}

\begin{abstract}
We explore effects of the light vector $U$-boson, which is weakly
coupled to nucleons, on the transition density $\rho _{t}$ and
pressure $P_{t}$ at the inner edge separating the liquid core from
the solid crust of neutron stars. Three methods, i.e., the
thermodynamical approach, the curvature matrix approach and the
Vlasov equation approach are used to determine the transition
density $\rho _{t}$ with the Skyrme effective nucleon-nucleon
interactions. We find that the $\rho _{t}$ and $P_{t}$ depend on not
only the ratio of coupling strength to mass squared of the $U$-boson
$g^{2}/\mu ^{2}$ but also its mass $\mu $ due to the finite range
interaction from the $U$-boson exchange. In particular, our results
indicate that the $\rho _{t}$ and $P_{t}$ are sensitive to both
$g^{2}/\mu ^{2}$ and $\mu $ if the $U$-boson mass $\mu $ is larger
than about $2$ MeV. Furthermore, we show that both $g^{2}/\mu ^{2}$
and $\mu $ can have significant influence on the mass-radius
relation and the crustal fraction of total moment of inertia of
neutron stars. In addition, we study the exchange term contribution
of the $U$-boson based on the density matrix expansion method, and
demonstrate that the exchange term effects on the nuclear matter
equation of state as well as the $\rho _{t}$ and $P_{t}$ are
generally negligible.
\end{abstract}

\maketitle

\section{Introduction}

The possible existence of a neutral weakly coupled light spin-$1$ gauge $U$%
-boson \cite{Fay80}, which is originated from supersymmetric extensions of
the standard model with an extra $U(1)$ symmetry, has recently attracted
much attention due to its multifaceted influences in particle physics,
nuclear physics, astrophysics and cosmology. For instance, the $U$-boson can
provide annihilation of light dark matter which can be responsible for the
excess flux of $511$ keV photons coming from the central region of our
Galaxy observed by the SPI/INTEGRAL satellite \cite%
{Jean03,Boe04a,Boe04b,Boe04c}. It is also proposed that the
$U$-boson can be mediator of the putative \textquotedblleft fifth
force" providing a possible mechanism for non-Newtonian gravity,
i.e., the violation of the inverse-square-law (ISL) of Newtonian
gravitational force at short distance \cite%
{Fuj71,Ark98,Fis99,Pea01,Hoy03,Lon03,Ade03,Uza03,Dec05,Rey05,Pok06,Kap07,Nes08,Kam08,Aza08,New09,Ger10,Luc10}%
. Thus far various upper limits on the deviation from the ISL have
been put forward down to femtometer range \cite{Pok06,Kam08,Ade03}.
Furthermore, the $U$-boson can involve a rich phenomenology in
particle physics and nuclear physics and may have observable effects
in particle decays \cite{Fay07,Zhu07,Che08,Fay09} and nucleon
scattering processes \cite{Bar75,Nes04,Pok06,Nes08,Kam08}, which can
also put limits on the $U$-boson properties. Studying properties of
the $U$-boson is thus important for understanding the relevant new
physics beyond the standard model.

Very recently, the effects of the $U$-boson on the nuclear matter equation
of state (EOS) and neutron star structure have been investigated \cite%
{Kri09,Wen09,Zha11,Wen11} and it is shown that the vector $U$-boson can
significantly stiffen the nuclear matter EOS\ and thus enhance drastically
the maximum mass of neutron stars. In particular, by considering the $U$%
-boson, the stability and observed global properties of neutron
stars can be reasonably explained by using the neutron-rich matter
EOS with a supersoft nuclear symmetry energy at supersaturation
densities consistent with the available terrestrial laboratory data
on the $\pi ^{-}/\pi ^{+}$ radio in relativistic heavy-ion
collisions from FOPI/GSI \cite{Rei07,Xia09}, while the supersoft
nuclear symmetry energy at supersaturation densities generally can
not support a canonical\ mass ($1.4M_{\odot }$) neutron star if the
$U$-boson is not introduced \cite{Wen09}. The $U$-boson has also
been introduced to
describe the recently discovered new holder of neutron star maximum mass of $%
1.97\pm 0.04$ $M_{\odot }$ from PSR J1614-2230 \cite{Dem10} using
soft EOS's consistent with existing terrestrial nuclear laboratory
experiments for hybrid neutron stars containing a quark core
described by MIT bag model using reasonable parameters
\cite{XuJ10a,Wen11}, and it is found that the constraints on the
$U$-boson properties are consistent with existing
constraints from neutron-proton and neutron-lead scatterings \cite%
{Nes08,Kam08} as well as the spectroscopy of antiproton atoms \cite{Pok06}.

In the studies about the $U$-boson influences on neutron star
structure~\cite{Kri09,Wen09,Zha11,Wen11}, the exchange term
contribution of the $U$-boson to the nuclear matter EOS has been
neglected and only the direct term contribution has been considered,
leading to that the nuclear matter EOS depends only on
the ratio the coupling strength to mass squared of the $U$-boson, namely, $%
g^{2}/\mu ^{2}$. Physically, the exchange term contribution of the
$U$-boson to the nuclear matter EOS\ will depend on both the
coupling constant $g$ and the $U$-boson mass $\mu $ due to the
finite-range interaction mediated by the $U$-boson. It is thus
interesting to see how the exchange term contribution of the
$U$-boson will influence the nuclear matter EOS. Furthermore,
neutron stars are expected to have a solid inner crust surrounding a
liquid core. Knowledge on properties of the crust plays an important
role in understanding many astrophysical
observations \cite%
{BPS71,BBP71,Pet95a,Pet95b,Lat00,Lat07,Ste05,Lin99,Hor04,Bur06,Owe05}.
The inner crust spans the region from the neutron drip-out point to
the inner edge separating the solid crust from the homogeneous
liquid core. While the neutron drip-out density $\rho
_{\mathrm{out}}$ is relatively well determined to be about $4\times
10^{11}$ g/cm$^{3}$ \cite{Rus06}, the transition density $\rho _{t}$
at the inner edge is still largely uncertain mainly because of our
very limited knowledge on the EOS of neutron-rich nucleonic matter,
especially the density dependence of the symmetry energy
\cite{Lat00,Lat07}. The transition density $\rho _{t}$ and the
corresponding pressure $P_{t}$ at the inner edge might be measurable
indirectly from observations of pulsar glitches \cite{Lat07,Lin99}.
Since the $U$-boson can have significant influence on the nuclear
matter EOS, it is therefore very interesting to see how the
$U$-boson will affect the inner edge of neutron star crusts.

In the present work, we investigate effects of the light vector $U$-boson on
the transition density $\rho _{t}$ and pressure $P_{t}$ at the inner edge of
neutron stars crust. The density matrix expansion (DME) approach \cite%
{Neg72,XuJ10b} is used to describe the exchange term contribution of
the finite-range interaction due to the $U$-boson exchange. Based on
the Skyrme effective nucleon-nucleon interactions, we use three
methods, i.e., the thermodynamical approach, the curvature matrix
approach and the Vlasov equation approach to determine the
transition density $\rho _{t}$. As expected, our results indicate
that the $\rho _{t}$ and $P_{t}$ depend on not only the ratio of
coupling strength to mass squared of the $U$-boson $g^{2}/\mu ^{2}$
but also its mass $\mu $ due to the finite range interaction from
the $U$-boson exchange. Furthermore, we find that the $\rho _{t}$
and $P_{t}$ are sensitive to both $g^{2}/\mu ^{2}$ and $\mu $ if the
$U$-boson mass $\mu $ is larger than about $2$ MeV and both
$g^{2}/\mu ^{2}$ and $\mu $ can have significant influence on the
mass-radius relation and the crustal fraction of total moment of
inertia of neutron stars. We also demonstrate that the exchange term
has minor influence on the nuclear matter EOS as well as the $\rho
_{t}$ and $P_{t}$ for the parameter values of $g^{2}/\mu ^{2}$ and
$\mu $ considered in this work.

\section{Theoretical models and methods}

\subsection{Nuclear matter symmetry energy}

The EOS of isospin asymmetric nuclear matter, defined by its binding
energy per nucleon, can be expanded to $2$nd-order in isospin
asymmetry $\delta $ as
\begin{equation}
E(\rho ,\delta )=E_{0}(\rho )+E_{\mathrm{sym}}(\rho )\delta ^{2}+O(\delta
^{4}),  \label{EOSANM}
\end{equation}%
where $\rho =\rho _{n}+\rho _{p}$ is the baryon density with $\rho _{n}$ and
$\rho _{p}$ denoting the neutron and proton densities, respectively; $\delta
=(\rho _{n}-\rho _{p})/(\rho _{p}+\rho _{n})$ is the isospin asymmetry; $%
E_{0}(\rho )=E(\rho ,\delta =0)$ is the binding energy per nucleon in
symmetric nuclear matter, and the nuclear symmetry energy is expressed as
\begin{equation}
E_{\mathrm{sym}}(\rho )=\frac{1}{2!}\frac{\partial ^{2}E(\rho ,\delta )}{%
\partial \delta ^{2}}|_{\delta =0}.  \label{Esym}
\end{equation}%
In Eq. (\ref{EOSANM}), the absence of odd-order terms in $\delta $
is due to the exchange symmetry between protons and neutrons in
nuclear matter when one neglects the Coulomb interaction and assumes
the charge symmetry of nuclear forces. The higher-order terms in
$\delta $ are negligible, leading to the well-known empirical
parabolic law for the EOS of asymmetric nuclear matter, which has
been verified by all many-body theories to date, at least for
densities up to moderate values (See, e.g., Ref. \cite{LCK08}). As a
good approximation, the density-dependent symmetry energy $E_{\mathrm{sym}%
}(\rho )$ can thus be extracted from $E_{\mathrm{sym}}(\rho )\approx E(\rho
,\delta =1)-E(\rho ,\delta =0)$.

Around the nuclear matter saturation density $\rho _{0}$, the nuclear
symmetry energy $E_{\mathrm{sym}}(\rho )$\ can be expanded as
\begin{equation}
E_{\mathrm{sym}}(\rho )=E_{\text{\textrm{sym}}}({\rho _{0}})+\frac{L}{3}(%
\frac{\rho -{\rho _{0}}}{{\rho _{0}}})+O((\frac{\rho -{\rho _{0}}}{{\rho _{0}%
}})^{2}),
\end{equation}%
where $L$ is the slope parameter of the nuclear symmetry energy at $\rho
_{0} $, i.e.,
\begin{equation}
L=3\rho _{0}\frac{\partial E_{\mathrm{sym}}(\rho )}{\partial \rho }|_{\rho
=\rho _{0}}.  \label{L}
\end{equation}%
The slope parameter $L$ characterizes the density dependence of the nuclear
symmetry energy around normal nuclear matter density, and thus carry
important information on the properties of nuclear symmetry energy at both
high and low densities.

The EOS of isospin asymmetric nuclear matter is a basic ingredient
to determine the properties of neutron stars. For symmetric nuclear
matter with equal fractions of neutrons and protons, its EOS
$E_{0}(\rho )$ is relatively well-determined. In particular, the
incompressibility $K_0$ of symmetric nuclear matter at its
saturation density $\rho _{0}$ has been determined to be $240\pm 20$
MeV from analyses of the nuclear giant monopole
resonances (GMR) \cite{You99,Lui04,Ma02,Vre03,Col04,Shl06,LiT07,Gar07,Col09,Che11b}%
, and its EOS at densities of $2\rho _{0}<\rho <5\rho _{0}$ has also
been constrained by measurements of collective flows \cite{Dan02a}
and subthreshold kaon production \cite{Aic85,Fuc06a} in relativistic
nucleus-nucleus collisions. On the other hand, the EOS of asymmetric
nuclear matter, especially the density dependence of the nuclear
symmetry energy, is largely unknown. Although the nuclear symmetry
energy at $\rho _{0}$ is known to be around $30$ MeV from the
empirical liquid-drop mass formula \cite{Mey66,Pom03}, its values at
other densities, especially at supra-saturation densities, are
poorly known \cite{Bar05,LCK08}. During the last decade, significant
progress has been made both experimentally and theoretically on
constraining the behavior of the symmetry energy at subsaturation
density and the value of $L$ constrained from different experimental
data or methods has become consistently convergent to about $60\pm
30$ MeV~\cite{Che05,Tsa09,Cen09} (See, e.g., Refs.
\cite{XuC10,Che11a} for recent summary). Furthermore, the IBUU04
transport model analysis of the FOPI data on the $\pi ^{-}/\pi ^{+}$
ratio in central heavy-ion collisions at SIS/GSI \cite{Rei07}
energies suggests a very soft symmetry energy at the supersaturation
densities~\cite{Xia09} (See also Refs. ~\cite{Fen10,Rus11}). These
studies have significantly improved our understanding for the EOS of
asymmetric nuclear matter, which have important implications on the
neutron star physics.

\subsection{Skyrme-Hartree-Fock approach}

For the nuclear effective interaction, we use in the present work
the so-called standard Skyrme force (see, e.g., Ref.~\cite{Cha97}),
which has been shown to be very successful in describing the
structure of finite nuclei. In the standard Skyrme Hartree-Fock
(SHF) approach, the nuclear effective interaction is taken to have a
zero-range, density- and momentum-dependent form \cite{Cha97}, i.e.,
\begin{eqnarray}
V_{12}(\mathbf{R},\mathbf{r}) &=&t_{0}(1+x_{0}P_{\sigma })\delta (\mathbf{r})
\notag \\
&+&\frac{1}{6}t_{3}(1+x_{3}P_{\sigma })\rho ^{\sigma }(\mathbf{R})\delta (%
\mathbf{r})  \notag \\
&+&\frac{1}{2}t_{1}(1+x_{1}P_{\sigma })(K^{^{\prime }2}\delta (\mathbf{r}%
)+\delta (\mathbf{r})K^{2})  \notag \\
&+&t_{2}(1+x_{2}P_{\sigma })\mathbf{K}^{^{\prime }}\cdot \delta (\mathbf{r})%
\mathbf{K}  \notag \\
&\mathbf{+}&iW_{0}(\mathbf{\sigma }_{1}+\mathbf{\sigma }_{2})\cdot \lbrack
\mathbf{K}^{^{\prime }}\times \delta (\mathbf{r})\mathbf{K]},  \label{V12Sky}
\end{eqnarray}%
with $\mathbf{r}={\vec{r}}_{1}-{\vec{r}}_{2}$ and $\mathbf{R}=(\vec{r}_{1}+%
\vec{r}_{2})/2$. In the above expression, the relative momentum operators $%
\mathbf{K}=(\mathbf{\nabla }_{1}-\mathbf{\nabla }_{2})/2i$ and $\mathbf{K}%
^{\prime }=-(\mathbf{\nabla }_{1}-\mathbf{\nabla }_{2})/2i$ act on the wave
function on the right and left, respectively. The quantities $P_{\sigma }$
and $\sigma _{i}$ denote, respectively, the spin exchange operator and Pauli
spin matrices. The $\sigma $, $t_{0}-t_{3}$, $x_{0}-x_{3}$ are the $9$
Skyrme interaction parameters and $W_{0}$ is the spin-orbit coupling
constant.

Within the standard form (see Eq.~(\ref{V12Sky})), the total energy of the
nuclear system can be written as:
\begin{equation}
E=\int {\mathcal{H}(\mathbf{r})d^{3}r},  \label{total_energy}
\end{equation}%
with $\mathcal{H}$ the Skyrme energy density. In the standard SHF model, the
total energy density of a spin-saturated nuclear system considered in this
work is written as~\cite{Cha97}
\begin{equation}
\mathcal{H}=\mathcal{K}+\mathcal{H}_{0}+\mathcal{H}_{3}+\mathcal{H}_{eff}+%
\mathcal{H}_{fin}+\mathcal{H}_{Coul}  \label{HSky}
\end{equation}%
where $\mathcal{K}=\frac{\hbar ^{2}}{2m}\tau $ is the kinetic-energy term
and $\mathcal{H}_{0}$, $\mathcal{H}_{3}$, $\mathcal{H}_{eff}$, $\mathcal{H}%
_{fin}$ are given by
\begin{eqnarray}
\mathcal{H}_{0} &=&t_{0}[(2+x_{0})\rho ^{2}-(2x_{0}+1)(\rho _{p}^{2}+\rho
_{n}^{2})]/4  \label{skyrme1} \\
\mathcal{H}_{3} &=&t_{3}\rho ^{\sigma }[(2+x_{3})\rho ^{2}-(2x_{3}+1)(\rho
_{p}^{2}+\rho _{n}^{2})]/24 \\
\mathcal{H}_{eff} &=&[t_{2}(2x_{2}+1)-t_{1}(2x_{1}+1)](\tau _{n}\rho
_{n}+\tau _{p}\rho _{p})/8  \notag \\
&&+[t_{1}(2+x_{1})+t_{2}(2+x_{2})]\tau \rho /8 \\
\mathcal{H}_{fin} &=&[3t_{1}(2+x_{1})-t_{2}(2+x_{2})](\nabla \rho )^{2}/32
\notag \\
&&-[3t_{1}(2x_{1}+1)+t_{2}(2x_{2}+1)]  \notag \\
&&\times \left[ (\nabla \rho _{n})^{2}+(\nabla \rho _{p})^{2}\right] /32
\label{Hfin0}
\end{eqnarray}%
in terms of the $9$ Skyrme interaction parameters $\sigma $, $t_{0}-t_{3}$, $%
x_{0}-x_{3}$. In the above equations, $\rho _{i}$ and $\tau _{i}$ are,
respectively, the local nucleon number and kinetic energy densities, whereas
$\rho $ and $\tau $ are corresponding total densities. $\mathcal{H}_{Coul}$
is the Coulomb term given by%
\begin{equation}
\mathcal{H}_{Coul}=\frac{1}{2}e^{2}\rho _{p}(\vec{r})\int \frac{\rho _{p}(%
\vec{r}^{\prime })}{\left\vert \vec{r}-\vec{r}^{\prime }\right\vert }d\vec{r}%
^{\prime }-\frac{3}{4}e^{2}\left( \frac{3}{\pi }\right) ^{1/3}\rho
_{p}^{4/3}(\vec{r}).
\end{equation}

The nucleon single-particle energy can be obtained from minimizing the total
energy of the nuclear system with respect to its wave function as
\begin{equation}
\epsilon _{q}=\frac{p^{2}}{2m}+U_{q}=\frac{p^{2}}{2m_{q}^{\star }}%
+U_{q}^{\star },~q=n,p,  \label{SPE}
\end{equation}%
where $U_{q}$ is the single particle potential while $m_{q}^{\star }$ and $%
U_{q}^{\star }$ represent, respectively, the nucleon effective mass
and effective single particle potential, which can be expressed,
respectively, as

\begin{eqnarray}
\frac{\hbar ^{2}}{2m_{q}^{\star }} &=&\frac{\hbar ^{2}}{2m_{q}}+\frac{1}{8}%
\rho \lbrack t_{1}(2+x_{1})+t_{2}(2+x_{2})]  \notag \\
&+&\frac{1}{8}\rho _{q}[t_{2}(2x_{2}+1)-t_{1}(2x_{1}+1)],  \label{EffMass}
\end{eqnarray}%
and
\begin{eqnarray}
U_{q}^{\star } &=&\frac{1}{2}t_{0}[(2+x_{0})\rho -(2x_{0}+1)\rho _{q}]
\notag \\
&+&\frac{1}{24}\sigma t_{3}\rho ^{\sigma -1}[(2+x_{3})\rho
^{2}-(2x_{3}+1)(\rho _{p}^{2}+\rho _{n}^{2})]  \notag \\
&+&\frac{1}{12}t_{3}\rho ^{\sigma }[(2+x_{3})\rho -(2x_{3}+1)\rho _{q}]
\notag \\
&+&\frac{1}{8}[t_{1}(2+x_{1})+t_{2}(2+x_{2})]\tau  \notag \\
&+&\frac{1}{8}[t_{2}(2x_{2}+1)-t_{1}(2x_{1}+1)]\tau _{q}  \notag \\
&+&\frac{1}{16}[t_{2}(2+x_{2})-3t_{1}(2+x_{1})]\nabla ^{2}\rho  \notag \\
&+&\frac{1}{16}[3t_{1}(2x_{1}+1)+t_{2}(2x_{2}+1)]\nabla {}^{2}\rho _{q}
\label{potential}
\end{eqnarray}%
For protons, the additional Coulomb potential is given by%
\begin{equation}
U_{Coul}=e^{2}\int \frac{\rho _{p}(\vec{r}^{\prime })}{\left\vert \vec{r}-%
\vec{r}^{\prime }\right\vert }d\vec{r}^{\prime }-e^{2}\left( \frac{3\rho
_{p}(\vec{r})}{\pi }\right) ^{1/3}.
\end{equation}

For the kinetic energy density, we use in this work the results from
the extended Thomas-Fermi approximation~\cite{Bra85}, i.e.,
\begin{equation}
\tau _{q}=a\rho _{q}^{5/3}+b\frac{(\nabla \rho _{q})^{2}}{\rho _{q}}+c\nabla
^{2}\rho _{q},  \label{etf}
\end{equation}%
where $a=\frac{3}{5}(3\pi ^{2})^{2/3}$, $b=1/36$ and $c=1/3$. From
the nucleon single-particle energy, the nucleon chemical potential
in infinite nuclear matter, i.e., $\mu _{q}$, can be obtained as the
value of the single-particle energy without gradient terms at the
Fermi surface $p_{q}^{F}=\hbar (3\pi ^{2}\rho _{q})^{1/3}$.

\subsection{The weakly coupled light vector $U$-boson}

Fujii \cite{Fuj71} first proposed that the non-Newtonian gravity can be
described by adding a Yukawa term to the conventional gravitational
potential between two objects of mass $m_{1}$ and $m_{2}$, i.e.,
\begin{equation}
V_{gra}(r)=-\frac{Gm_{1}m_{2}}{r}(1+\alpha e^{-r/\lambda }),  \label{Vgra}
\end{equation}%
where $\alpha $ is a dimensionless strength parameter, $\lambda $ is
the length scale and $G$ is the gravitational constant. In the boson
exchange picture, the light and weakly coupled vector $U$-boson is a
favorite candidate mediating the extra interaction for the
non-Newtonian gravity \cite{Fay80}, leading to the finite-range
Yukawa potential between two nucleons which can be expressed as
\begin{equation}
V_{\mathrm{UB}}(r)=\frac{g^{2}}{4\pi }\frac{e^{-\mu r}}{r},  \label{VYuk}
\end{equation}%
where $g$ and $\mu $ represent the $U$-boson-nucleon coupling
constant and the $U$-boson mass, respectively. Comparing
Eq.~(\ref{VYuk}) with the Yukawa term in Eq.~(\ref{Vgra}), one can
find the relations $\alpha =-g^{2}/(4\pi Gm^{2})$ and $\lambda
=1/\mu $ (in natural units) where $m$ is the nucleon mass. By adding
the Yukawa potential of Eq.~(\ref{VYuk}) to the standard Skyrme
effective nucleon-nucleon interaction in Eq.~(\ref{V12Sky}), the
extra binding energy of the nuclear system due to the $U$-boson can
be expressed as the integral of energy density
$\mathcal{H}_{\mathrm{UB}}$ as follows
\begin{equation}
E_{\mathrm{UB}}=\int \mathcal{H}_{\mathrm{UB}}(\boldsymbol{r})d^{3}r,
\label{EUB}
\end{equation}%
with
\begin{equation}
\mathcal{H}_{\mathrm{UB}}=\mathcal{H}_{\mathrm{UB}}^{D}+\mathcal{H}_{\mathrm{%
UB}}^{E},  \label{HUB}
\end{equation}%
where $\mathcal{H}_{\mathrm{UB}}^{D}$ and
$\mathcal{H}_{\mathrm{UB}}^{E}$ are the direct and exchange
contribution to the energy density, respectively. For the
finite-range Yukawa interaction of Eq.~(\ref{VYuk}), the direct term
contribution to the energy density can be easily obtained as
\cite{Kri09,Wen09,Zha11,Wen11}
\begin{equation}
\mathcal{H}_{\mathrm{UB}}^{D}=\frac{1}{2V}\int \rho (\vec{r}_{1})\frac{g^{2}%
}{4\pi }\frac{e^{-\mu r}}{r}\rho (\vec{r}_{2})d\vec{r}_{1}d\vec{r}_{2}=\frac{%
1}{2}\frac{g^{2}}{\mu ^{2}}\rho ^{2},  \label{HUBD}
\end{equation}%
where $V$ is the normalization volume, $\rho =\rho _{n}+\rho _{p}$ is the
baryon number density, and $r=|\vec{r}_{1}-\vec{r}_{2}|$.

Although the direct term contribution of a finite-range interaction
to the nuclear energy density can be treated exactly, it is
numerically challenging to evaluate the exchange contribution. The
latter can be, however, approximated by that from a Skyrme-like
zero-range interaction using the density-matrix expansion
\cite{Neg72,XuJ10b}, and the results can be
obtained as%
\begin{eqnarray}
\mathcal{H}_{\mathrm{UB}}^{E} &=&\sum_{q=p,n}\frac{g^{2}}{4}\left[ 2\rho
_{q}\tau _{q}I_{1q}+\frac{1}{2}\left( I_{1q}+\rho _{q}\frac{\partial {I_{1q}}%
}{\partial {\rho _{q}}}\right) (\nabla {\rho _{q}})^{2}\right]  \notag \\
&-&\sum_{q=p,n}\frac{g^{2}}{4}\left[ \frac{6}{5}(3\pi ^{2})^{2/3}\rho
_{q}^{8/3}I_{1q}+\rho _{q}^{2}I_{2q}\right] .  \label{HUBE}
\end{eqnarray}%
The integrations in Eq.~(\ref{HUBE}) are defined as
\begin{eqnarray}
I_{1q} &=&\int {drr^{4}\rho _{SL}(k_{q}^{F}r)g(k_{q}^{F}r)\frac{e^{-\mu {r}}%
}{r},} \\
I_{2q} &=&\int {drr^{2}\rho _{SL}^{2}(k_{q}^{F}r)\frac{e^{-\mu {r}}}{r}},
\end{eqnarray}%
with
\begin{eqnarray}
\rho _{SL}(k_{q}^{F}r) &=&\frac{3}{k_{q}^{F}r}j_{1}(k_{q}^{F}r), \\
g(k_{q}^{F}r) &=&\frac{35}{2(k_{q}^{F}r)^{3}}j_{3}(k_{q}^{F}r),
\end{eqnarray}%
where $j_{1}$ and $j_{3}$ are, respectively, the first- and third-order
spherical Bessel functions and $k_{q}^{F}=(3\pi ^{2}\rho _{q})^{1/3}$ is the
Fermi momentum.

One can see from Eq.~(\ref{HUBD}) that for the direct term, the
$U$-boson contributes to the nuclear energy density only through the
combination $g^{2}/\mu ^{2}$. On the other hand, it is indicated
from Eq.~(\ref{HUBE}) that the exchange term contribution to the
nuclear energy density depends on both the coupling constant $g$ and
the mass $\mu $ in a complicated way. Furthermore, the density
gradient terms appear automatically in the exchange term
contribution to the nuclear energy density due to the density matrix
expansion of the finite-range Yukawa potential. It is interesting to
see that the exchange term contribution to the nuclear matter EOS
further depends on the isospin asymmetry although the direct term
contribution to the nuclear matter EOS is isospin independent due to
the fact that the $U$-boson is an isoscalar boson. Therefore, the
$U$-boson will contribute to the nuclear symmetry energy through the
exchange term. However, as will be shown later, the exchange term
contribution to the nuclear matter EOS and the symmetry energy is
quite small and can be neglected safely within the parameter value
region of the coupling constant $g$ and the $U$-boson mass $\mu $
considered in the present work.

The nucleon single-particle energy due to the $U$-boson can be
obtained from variation of the corresponding energy of
Eq.~(\ref{EUB}) with respect to its wave function. The $U$-boson
contribution to the nucleon effective mass can be expressed as
\begin{equation}
\frac{\hbar ^{2}}{2m_{q,\mathrm{UB}}^{\star }}=\frac{g^{2}}{2}\rho
_{q}I_{1q},  \label{EffMassUB}
\end{equation}%
which should be added to the right hand side of Eq.~(\ref{EffMass})
to obtain the total nucleon effective mass $m_{q}^{\star }$. It
should be noted that the $U$-boson contribution to the nucleon
effective mass is from the exchange term with the density matrix
expansion of the finite-range Yukawa potential, which leads to the
isospin dependent contribution to the nucleon effective mass
although the $U$-boson is isoscalar. The $U$-boson contribution to
the effective single-particle potential can be written as

\begin{eqnarray}
U_{q,\mathrm{UB}}^{\star } &=&\frac{g^{2}}{4\pi }\int \frac{e^{-\mu
\left\vert \vec{r}-\vec{r}^{\prime }\right\vert }}{\left\vert \vec{r}-\vec{r}%
^{\prime }\right\vert }\rho (\vec{r}^{\prime })d\vec{r}^{\prime }  \notag \\
&-&\frac{g^{2}}{4}\left[ \rho _{q}\left( 2I_{2q}+\rho_{q}\frac{\partial {I_{2q}}}{%
\partial {\rho _{q}}}\right) +2(3\pi ^{2})^{2/3}\rho _{q}^{5/3}I_{1q}\right]
\notag \\
&+&\frac{g^{2}}{4}\left( \frac{1}{18}\frac{I_{1q}}{\rho _{q}}-\frac{17}{18}%
\frac{\partial I_{1q}}{\partial \rho _{q}}-\frac{1}{2}\rho _{q}\frac{%
\partial ^{2}{I_{1q}}}{\partial {\rho _{q}}^{{2}}}\right) (\nabla {\rho _{q}}%
)^{2}  \notag \\
&-&\frac{g^{2}}{12}\left( I_{1q}+\rho _{q}\frac{\partial {I_{1q}}}{\partial {%
\rho _{q}}}\right) \nabla ^{2}{\rho _{q}.}  \label{UUB}
\end{eqnarray}%
where the first term in the right hand side of Eq. (\ref{UUB}) is
from the direct term contribution while the other terms are from the
exchange term contribution. As expected, the direct term
contribution is isospin independent while the exchange term
contribution depends on the isospin asymmetry as well as the density
gradients. Furthermore, one can see that the $U$-boson contribution
to the single-particle potential depends on both the coupling
constant $g$ and the mass $\mu $ in a complicated way.

\subsection{The transition density in neutron stars}

The transition density is the baryon number density that separates
the liquid core from the inner crust in neutron stars and it plays
an important role in determining the structural properties of
neutron stars such as the crustal fraction of total moment of
inertia and the mass-radius relations of static neutron stars. In
principle, the transition density can be obtained from comparing
relevant properties of the nonuniform solid crust and the uniform
liquid core mainly consisting of neutrons, protons, and electrons
(\emph{npe} matter). However, this is practically very difficult
since the inner crust may contain the so-called \textquotedblleft
nuclear pasta\textquotedblright\ with very complicated
geometries~\cite{Lat04,Rav83,Oya93,Hor04,Ste08}. In practice, a good
approximation is to search for the density at which the uniform
liquid first becomes unstable against small amplitude density
fluctuations with clusterization. This approximation has been shown
to produce a very small error for the actual core-crust transition
density and it would yield the exact transition density for a
second-order phase transition \cite{Pet95b,Dou00,Dou01,Hor03}. So
far, several such methods including the thermodynamical
method~\cite{Kub07,Lat07,Wor08}, the dynamical curvature matrix
method \cite{BPS71,BBP71,Pet95a,Pet95b,Dou00,Oya07,Duc07}, the
Vlasov equation method \cite{Cho04,Pro06,Duc08a,Duc08b,Pai10}, and
the random phase approximation (RPA)~\cite{Hor01,Hor03,Duc08a} have
been applied extensively in the literature. Here, we briefly
introduce the thermodynamical method, the dynamical curvature matrix
method and the Vlasov equation method, which will be used to
calculate the transition density in this work.

\subsubsection{The thermodynamical method for transition density in neutron
stars}

In the thermodynamical method, the system is required to obey the
following intrinsic stability condition~\cite{Cal85,Kub07,Lat07}
\begin{eqnarray}
-\left( \frac{\partial P}{\partial v}\right) _{\mu_{np} } &>&0,  \label{ther1} \\
-\left( \frac{\partial \mu_{np} }{\partial q_{c}}\right) _{v} &>&0,
\label{ther2}
\end{eqnarray}%
where the $P=P_{b}+P_{e}$ is the total pressure of the $npe$ matter
system with $P_{b}$ and $P_{e}$ denoting the contributions from
baryons and electrons respectively, and the $v$ and $q_{c}$ are the
volume and charge per baryon number. The $\mu_{np} $ is defined as
the chemical potential difference between neutrons and protons,
i.e.,
\begin{equation}
\mu_{np} =\mu _{n}-\mu _{p}.
\end{equation}%
The conditions of Eq.~(\ref{ther1}) and Eq.~(\ref{ther2}) are equivalent to
require the convexity of the energy per particle in the single phase~\cite%
{Kub07,Lat07} by ignoring the finite size effects due to surface and
Coulomb energies \cite{XuJ09}. In fact, Eq.~(\ref{ther1}) is simply
the well-known mechanical stability condition of the system at a
fixed $\mu_{np} $, which ensures that any local density fluctuation
will not diverge. On the other hand, Eq.~(\ref{ther2}) is the charge
or chemical stability condition of the system at a fixed density. It
means that any local charge variation violating the charge
neutrality condition will not diverge.

The pressure $P_{e}$ is only a function of the chemical potential
difference $\mu_{np} $ by assuming the $\beta $-equilibrium
condition is satisfied, i.e., $\mu_{np} =\mu _{e}$. By using the
relation $\frac{\partial E_{b}(\rho ,x_{p})}{\partial
x_{p}}=-\mu_{np} $ with $E_{b}(\rho ,x_{p})$ being energy per
baryon from the baryons in the $\beta $-equilibrium neutron star matter and $%
x_{p}=\rho _{p}/\rho $, and treating the electrons as free Fermi
gas, one can show \cite{XuJ09} that the thermodynamical relations
Eq.~(\ref{ther1}) and Eq.~(\ref{ther2}) are actually equivalent to
the following condition
\begin{eqnarray}
V_{ther} &=&2\rho \frac{\partial E_{b}(\rho ,x_{p})}{\partial \rho }+\rho
^{2}\frac{\partial ^{2}E_{b}(\rho ,x_{p})}{\partial \rho ^{2}}  \notag \\
&-&\left( \frac{\partial ^{2}E_{b}(\rho ,x_{p})}{\partial \rho \partial x_{p}%
}\rho \right) ^{2}/\frac{\partial ^{2}E_{b}(\rho ,x_{p})}{\partial x_{p}^{2}}%
>0,  \label{Vther}
\end{eqnarray}%
which determines the thermodynamical instability region of the $\beta $%
-equilibrium neutron star matter. The baryon number density that
violates the condition Eq.~(\ref{Vther}) then corresponds to the
core-crust transition density in neutron stars for the
thermodynamical method.

\subsubsection{The curvature matrix method for transition density in neutron
stars}

In the curvature matrix method, the instability region of
homogeneous nuclear matters against clusterization is determined by
introducing a finite-size spatially periodic density fluctuation
$\delta \rho $ to the system and then examining how the system free
energy varies with the fluctuation~\cite{Duc07}. The fluctuation
will affect the three components of homogeneous nuclear matter
(neutrons, protons and electrons) independently when assuming it
occurs only on finite microscopic scale in the $\beta $-equilibrium
nuclear matter as
\begin{equation}
\rho _{q}=\rho _{q}^{0}+\delta \rho _{q},  \label{fluctuation}
\end{equation}%
with $q=n$, $p$, $e$. Then the free energy $f$ at each point of
density $\rho _{q}(\boldsymbol{r})=\rho _{q}^{0}+\delta \rho _{q}$
can be expressed as
\begin{widetext}
\begin{eqnarray}
f(\rho _{q}) &=&f(\rho _{q}^{0})+\sum_{q=n,p,e}\left( \frac{\partial {f}}{%
\partial {\rho _{q}}}\right) _{0}\delta \rho _{q}  \notag \\
&+&\sum_{q,q^{\prime }=n,p,e}\frac{1}{2}\left( \frac{\partial ^{2}f}{%
\partial {\rho _{q}}\partial {\rho _{q^{\prime }}}}\right) _{0}\delta \rho
_{q}\delta \rho _{q^{\prime }}+\cdots .  \label{free_energy}
\end{eqnarray}%
For a stable homogeneous $npe$ matter system, the first-order term $\left(
\frac{\partial {f}}{\partial {\rho _{q}}}\right) _{0}$ in Eq.~(\ref%
{free_energy}) must equal to zero and a density fluctuation $\delta
\rho _{q} $ should lead to an increasing of the free energy, which
is equivalent to require the second-order term in
Eq.~(\ref{free_energy}) to be positive for any density fluctuation
$\delta \rho _{q}$. This can be ensured by the positive definiteness
of the following curvature matrix
\begin{eqnarray}
C_{CM}^{f} &=&\left(
\begin{array}{ccc}
\frac{\partial ^{2}f}{\partial \rho _{n}^{2}} & \frac{\partial ^{2}f}{%
\partial \rho _{n}\partial \rho _{p}} & \frac{\partial ^{2}f}{\partial \rho
_{n}\partial \rho _{e}} \\
\frac{\partial ^{2}f}{\partial \rho _{p}\partial \rho _{n}} & \frac{\partial
^{2}f}{\partial \rho _{p}^{2}} & \frac{\partial ^{2}f}{\partial \rho
_{p}\partial \rho _{e}} \\
\frac{\partial ^{2}f}{\partial \rho _{e}\partial \rho _{n}} & \frac{\partial
^{2}f}{\partial \rho _{e}\partial \rho _{p}} & \frac{\partial ^{2}f}{%
\partial \rho _{e}^{2}}%
\end{array}%
\right)  \notag \\
&=&\left(
\begin{array}{ccc}
\frac{\partial {\mu _{n}}}{\partial \rho _{n}} & \frac{\partial {\mu _{n}}}{%
\partial \rho _{p}} & 0 \\
\frac{\partial {\mu _{p}}}{\partial \rho _{n}} & \frac{\partial {\mu _{p}}}{%
\partial \rho _{p}} & 0 \\
0 & 0 & \frac{\partial {\mu _{e}}}{\partial \rho _{e}}%
\end{array}%
\right)
+ k^2 \left(
\begin{array}{ccc}
D _{nn} & D _{np} & 0 \\
D _{pn} & D _{pp} & 0 \\
0 & 0 & 0
\end{array}
\right)
+ \frac{g^2}{k^2+\mu^2} \left(
\begin{array}{ccc}
1 & 1 & 0 \\
1 & 1 & 0 \\
0 & 0 & 0
\end{array}
\right)
+ \frac{4\pi e^2}{k^2} \left(
\begin{array}{ccc}
0 & 0 & 0 \\
0 & 1 & -1 \\
0 & -1 & 1
\end{array}
\right) ,  \label{cur_matrix}
\end{eqnarray}
\end{widetext}%
where $k$ is the wave vector of the spatially periodic density
fluctuations and the effective density-gradient coefficients are
defined as
\begin{eqnarray}
D_{nn}(D_{pp}) &=& \frac{3}{16} [t_{1}(1-x_{1})-t_{2}(1+x_{2})]\notag\\
&-& \frac{1}{24} [t_{1}(1-x_{1})+3t_{2}(1+x_{2})]\notag\\
&+& \frac{g^{2}}{12}\left( I_{1n(p)}+\rho _{n(p)}\frac{\partial {I_{1n(p)}}}{\partial {\rho _{n(p)}}}\right),\\
D_{np}(D_{pn}) &=& \frac{1}{16} [3t_{1}(2+x_{1})-t_{2}(2+x_{2})]\notag\\
&-& \frac{1}{24} [t_{1}(2+x_{1})+t_{2}(2+x_{2})].
\end{eqnarray}
From Eq.~(\ref{cur_matrix}), one can see that the matrix
$C_{CM}^{f}$ includes four parts. The bulk part, i.e., the first
term in the right hand side of Eq.~(\ref{cur_matrix}) is $k$
independent but the following $3$ parts are all $k$ dependent due to
the density-gradient terms in Eq.~(\ref{potential}), the direct term
contribution of the finite range interaction from the $U$-boson
exchange, and the Coulomb interaction, respectively.

The matrix $C_{CM}^{f}$ is defined for each point ($\rho _{n}$,$\rho
_{p}$,$\rho _{e}$), and the sign of its three eigenvalues determines
the sign of the second-order term in Eq.~(\ref{free_energy}),
namely, only if all the eigenvalues of the matrix are positive, the
free energy of the system will remain the minimum value and the
nuclear system will be stable for all the density fluctuations. So
the baryon number density violating the positive definiteness of the
matrix $C_{CM}^{f}$ corresponds to the core-crust transition density
in neutron stars for the curvature matrix method.

\subsubsection{The Vlasov equation method for transition density in neutron
stars}

To determine the core-crust transition density in neutron stars
within the Vlasov equation method, we include here for completeness
a brief description for the method (See, e.g., Ref.\
\cite{XuJ10b,Cho04} for the details). For a $\beta $-stable and
electrically neutral $npe$ matter, the Vlasov equation can be
expressed as
\begin{equation}
\frac{\partial f_{q}(\vec{R},\vec{p},t)}{\partial t}+\vec{v}_{q}\cdot \nabla
_{\vec{R}}f_{q}(\vec{R},\vec{p},t)-\nabla _{\vec{R}}U_{q}\cdot \nabla _{\vec{%
p}}f_{q}(\vec{R},\vec{p},t)=0  \label{vlasov}
\end{equation}%
in terms of the semi-classical Wigner function for particle type
$q=n,p,e$, i.e.,
\begin{eqnarray}
f_{q}(\vec{R},\vec{p},t) &=&\frac{1}{(2\pi )^{3}}\sum_{i}\int \phi
_{qi}\left( \vec{R}-\frac{\vec{r}}{2},t\right)  \notag \\
&\times &\phi _{qi}^{\star }\left( \vec{R}+\frac{\vec{r}}{2},t\right) e^{i%
\vec{p}\cdot \vec{r}}d^{3}r,
\end{eqnarray}%
where $\phi _{qi}$ is the wave function of $i$-th particle of type $q$, $%
\vec{R}$ and $\vec{r}$ are defined as the same as $\mathbf{R}$ and
$\mathbf{r} $ in the previous section. In Eq.~(\ref{vlasov}),
$\vec{v}_{q}=\vec{p}/{(m_{q}^{\star }}^{2}+p^{2})^{1/2}$ denotes the
particle velocity and $m_{e}^{\star }=m_{e}$.

The density fluctuation due to a collective mode with frequency $%
\omega$ and wavevector $\vec{k}$ in nuclear matter can be studied
through following the standard procedure~\cite{Cho04} by writing
\begin{equation}
f_q(\vec{R},\vec{p},t) = f_q^0(\vec{p}) + \delta f_q (\vec{R},\vec{p},t)
\end{equation}
with
\begin{equation}
\delta f_q (\vec{R},\vec{p},t) = \delta \tilde{f_q} (\vec{p}) e^{-i\omega
t+i \vec{k}\cdot\vec{R}}.
\end{equation}
By expressing $\rho_q(\vec{R},t) = \rho_q^0 + \delta\rho_q
(\vec{R},t)$ with
\begin{equation}  \label{vlasov2}
\delta \rho_q(\vec{R},t) = \frac{2}{(2\pi)^3} \int \delta f_q(\vec{R},\vec{p}%
,t) d^3p,
\end{equation}
we obtain the Vlasov equation
\begin{eqnarray}  \label{drho}
\delta \rho_q &\approx& X_{q} L_{q} \left( \sum_{q^\prime} \frac{\delta U_q}{\delta \rho_{q^\prime}}
\delta \rho_{q^\prime}\right),
\end{eqnarray}
where $L_q$ is the usual Lindhard function
\begin{eqnarray}
L_q = \int_{-1}^{1} \frac{\cos\theta d(\cos\theta)}{s_q-\cos\theta}
= -2 + s_q \ln \left( \frac{s_q+1}{s_q-1}\right),
\end{eqnarray}
with $s_q=\omega/k v_q^F$ and $v_q^F =p_q^F/
({m_q^\star}^2+{p_q^F}^2)^{1/2}$ being the Fermi velocity. The
momentum integration can be evaluated approximately as
\begin{eqnarray}  \label{xq}
X_q &=& \frac{1}{2\pi^2} \int_0^{p_q^F} \left(-\frac{\partial f_q^0}{%
\partial\epsilon_q}\right) p^2 dp  \approx \frac{p_q^F m_q^\star}{2\pi^2}
\end{eqnarray}
for $q=n,p$ with $\epsilon_q^F \approx{p_q^F}^2/2m_q^\star$ and
\begin{equation}  \label{xe}
X_e \approx \frac{\mu_e^2}{2\pi^2},
\end{equation}
for electrons, where $\mu_e \approx p_e^F$ is the electron chemical
potential.

For protons, there are additional direct and exchange Coulomb contributions
to the factor $\sum_{q^\prime} \frac{\delta U_q}{\delta \rho_{q^\prime}}
\delta \rho_{q^\prime}$ in Eq.~(\ref{drho}) given, respectively, by
\begin{eqnarray}
\delta U_p^{CD} &=& \frac{4\pi e^2}{k^2} (\delta \rho_p - \delta \rho_e), \\
\delta U_p^{CE} &=& -\frac{1}{3} e^2 \left( \frac{3}{\pi} \right)^{1/3}
\rho_p^{-2/3} \delta \rho_p.
\end{eqnarray}
For electrons, there are only direct and exchange Coulomb contributions to $%
\sum_{q^\prime} \frac{\delta U_q}{\delta \rho_{q^\prime}} \delta
\rho_{q^\prime}$.

After linearizing the Vlasov equation, we can reexpress
Eq.~(\ref{drho}) as a function of the collective density fluctuation
\begin{equation}
C_{VE}^{f}(\delta \rho _{n},\delta \rho _{p},\delta \rho
_{e})^{T}=0, \label{eqdrho}
\end{equation}%
with%
\begin{widetext}
\begin{eqnarray}
C_{VE}^{f}&=&\left(
\begin{array}{ccc}
X_{n}L_{n}\frac{\partial U_{n}}{\partial \rho _{n}}-1 & X_{n}L_{n}\frac{%
\partial U_{n}}{\partial \rho _{p}} & 0 \\
X_{p}L_{p}\frac{\partial U_{p}}{\partial \rho _{n}} & X_{p}L_{p}\frac{%
\partial U_{p}}{\partial \rho _{p}}-1 & X_{p}L_{p}\frac{\partial U_{p}}{%
\partial \rho _{e}} \\
0 & X_{e}L_{e}\frac{\partial U_{e}}{\partial \rho _{p}} & X_{e}L_{e}\frac{%
\partial U_{e}}{\partial \rho _{e}}-1%
\end{array}%
\right)\notag\\
&=& \left(
\begin{array}{ccc}
X_{n}L_{n} & 0 & 0 \\
0 & X_{p}L_{p} & 0 \\
0 & 0 & X_{e}L_{e}
\end{array}
\right)
\left(
\begin{array}{ccc}
\frac{\partial U_{n}}{\partial \rho _{n}} & \frac{\partial U_{n}}{\partial \rho _{p}} & 0 \\
\frac{\partial U_{p}}{\partial \rho _{n}} & \frac{\partial U_{p}}{\partial \rho _{p}} & \frac{\partial U_{p}}{\partial \rho _{e}} \\
0 & \frac{\partial U_{e}}{\partial \rho _{p}} & \frac{\partial U_{e}}{\partial \rho _{e}}
\end{array}
\right)
- \left(
\begin{array}{ccc}
1 & 0 & 0 \\
0 & 1 & 0 \\
0 & 0 & 1
\end{array}
\right) .  \label{vla_matrix}
\end{eqnarray}
\end{widetext}
In Eq.~(\ref{vla_matrix}), the $U_q$ is the single-particle
potential in the \emph{npe} system and the matrix
\begin{eqnarray}
U=\left(
\begin{array}{ccc}
\frac{\partial U_{n}}{\partial \rho _{n}} & \frac{\partial U_{n}}{\partial \rho _{p}} & 0 \\
\frac{\partial U_{p}}{\partial \rho _{n}} & \frac{\partial U_{p}}{\partial \rho _{p}} & \frac{\partial U_{p}}{\partial \rho _{e}} \\
0 & \frac{\partial U_{e}}{\partial \rho _{p}} & \frac{\partial U_{e}}{\partial \rho _{e}}
\end{array}
\right)
\end{eqnarray}
has the same $k$-dependent terms as in Eq.~(\ref{cur_matrix}).

The determinant $|C_{VE}^{f}|=0$ determines the dispersion relation
$\omega (k)$ of the collective density fluctuation which also
determines the non-trivial solutions of Eq.~(\ref{eqdrho}). The
transition density in neutron stars is the density at which the
frequency $\omega $ becomes imaginary, leading to that the
collective density fluctuation would grow exponentially and thus the
instability of the neutron star matter occur. To determine the
condition for this to occur, we let $s_{q}=-i\nu _{q}$ $(\nu
_{q}>0)$ and rewrite the Lindhard function as $L_{q}=-2+2\nu
_{q}\arctan (1/\nu _{q})$. Since the values of $L_{q}$ are in the
range of $-2<L_{q}<0$, the critical values $L_{n}=L_{p}=L_{e}=-2 $,
corresponding to $\nu _{q}=0$, then determine the spinodal boundary
of the system when they are substituted into $|C_{VE}^{f}|=0$. The
baryon number density that makes $|C_{VE}^{f}|$ vanish then
corresponds to the spinodal boundary in the neutron star matter or
the core-crust transition density of neutron stars for the Vlasov
equation method.

\section{Results}

In the present work, for the Skyrme effective nucleon-nucleon
interaction, we use the modified Skyrme-like (MSL)
parameter~\cite{Che10,Che11} for which the $9$ Skyrme interaction
parameters $\sigma $, $t_{0}-t_{3}$, $x_{0}-x_{3}$ are obtained
analytically in terms of $9$ macroscopic quantities $\rho _{0}$,
$E_{0}(\rho _{0})$, the incompressibility $K_{0}$, the isoscalar
effective mass $m_{s,0}^{\ast }$,
the isovector effective mass $m_{v,0}^{\ast }$, $E_{\text{\textrm{sym}}}({%
\rho _{0}})$, $L$, the gradient coefficient $G_{S}$, and the
symmetry-gradient coefficient $G_{V}$. In particular, the MSL0
parameter set~\cite{Che10} is obtained by using the following
empirical values for the $9$ macroscopic quantities: $\rho
_{0}=0.16$ fm$^{-3}$, $E_{0}(\rho _{0})=-16$ MeV, $K_{0}=230$ MeV,
$m_{s,0}^{\ast }=0.8m$, $m_{v,0}^{\ast }=0.7m$,
$E_{\text{\textrm{sym}}}({\rho _{0}})=30$ MeV, $L=60$ MeV, $G_{V}=5$
MeV$\cdot $fm$^{5}$, and $G_{S}=132$ MeV$\cdot $fm$^{5}$. And the
spin-orbit coupling constant $W_{0}=133.3$ MeV $\cdot $fm$^{5}$ is
used to fit the neutron $p_{1/2}-p_{3/2}$ splitting in $^{16}$O. It
has been shown~\cite{Che10} that the MSL0 interaction can give a
good description of the binding energies and charge rms radii for a
number of closed-shell or semi-closed-shell nuclei.

For the $U$-boson-nucleon coupling constant $g$ and the $U$-boson
mass $\mu $, their values are largely uncertain. As argued by
Krivoruchenko \textsl{et al.} \cite{Kri09}, in order to ensure that
the $U$-boson effects on finite nuclei should be negligible, the
Compton wavelength of the $U$-boson is usually assumed to be greater
than the radius of heavy nuclei, i.e., about $7$ fm, leading to $\mu
\lesssim 30$ MeV. At the same time, the $g^{2}/\mu ^{2}$ value of
the $U$-boson should be less than about $200$ GeV$^{-2}$, which
roughly corresponds to the value of the ordinary vector $\omega $
boson. Otherwise, the $U$-boson is neither weakly coupled nor light.
On the other hand, there also exist some constraints on properties
of the $U$-boson from cosmology and astrophysical observations. For
example, the $U$-boson mass $\mu $ is required to exceed the mass of
light cold dark matter, i.e., $\sim $ MeV, to explain the excess
flux of $511$ keV photons coming from the central region of our
Galaxy observed by the SPI/INTEGRAL satellite \cite{Jac07}. Based on
above discussions, in the present work we assume the $U$-boson mass
is in the range of $2$ MeV $\lesssim \mu \lesssim 30$ MeV (the
corresponding Compton wavelength of the $U$-boson is thus between
about $7$ fm and $100$ fm) and the $g^{2}/\mu ^{2}$ value of the
$U$-boson satisfies $g^{2}/\mu ^{2}\lesssim 150$ GeV$^{-2}$. The
latter is further consistent with existing constraints from
neutron-proton and neutron-lead scatterings, the spectroscopy of
antiproton atoms as well as the recently discovered new holder of
neutron star maximum mass of $1.97\pm 0.04$ $M_{\odot }$ from PSR
J1614-2230 \cite{Nes08,Kam08,Pok06,Wen11}.

\subsection{Nuclear matter symmetry energy from the $U$-boson}

As have been seen in the previous section, the direct term of the
$U$-boson contributes to the nuclear matter EOS only through the
combination $g^{2}/\mu ^{2}$, and in particular, its contribution to
the energy per nucleon is given by $\frac{1}{2}g^{2}/\mu ^{2}\rho $.
As it was emphasized by Fujii \cite{Fuj88}, for the direct term
contribution, though both the coupling constant $g$ and the mass
$\mu $ are small for the light and weakly coupled bosons, the value
of the ratio $g^{2}/\mu ^{2}$ can be large. Therefore, the light and
weakly coupled bosons can significantly affect the nuclear matter
EOS and thus the properties of neutron stars
\cite{Kri09,Wen09,Zha11,Wen11}. On the other hand, the exchange term
contribution to the nuclear matter EOS depends on both the coupling
constant $g$ and the mass $\mu $ in a complicated way. Furthermore,
it is interesting to see that the isoscalar $U$-boson will
contribute to the nuclear matter symmetry energy due to the exchange
term contribution though the direct term does not have such
contribution.
\begin{figure}[tbp]
\includegraphics[scale=1.0]{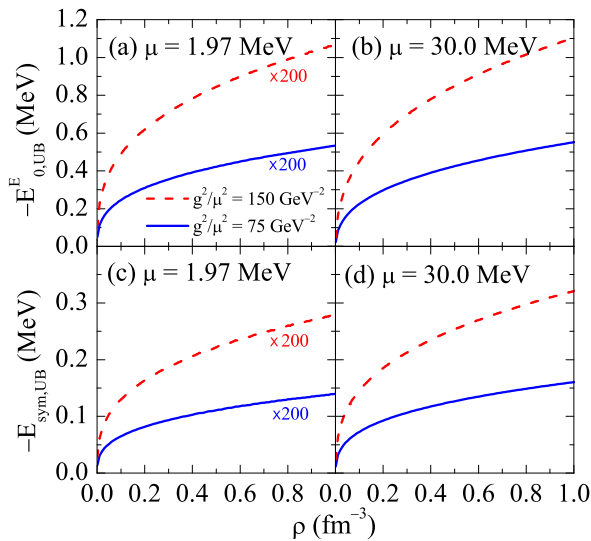}
\caption{(Color online) The exchange term contribution of the
$U$-boson to the EOS of symmetric nuclear matter ((a) and (b)) and
the symmetry energy ((c) and (d)) as functions of density with
several typical values of $\protect\mu $ and $g^{2}/\protect\mu
^{2}$. Note: The results with $\protect\mu =1.97$ MeV ((a) and (c))
have been rescaled by multiplying a factor of $200$ for
convenience.} \label{EsymUB}
\end{figure}

To see quantitatively how the exchange term of the $U$-boson affects the
nuclear matter EOS, we show in\ Fig. \ref{EsymUB} the exchange term
contribution of the $U$-boson to the EOS of symmetric nuclear matter $E_{%
\mathrm{0,UB}}^{E}(\rho )$ ((a) and (b)) and the nuclear matter
symmetry energy $E_{\mathrm{sym,UB}}(\rho )$ ((c) and (d)) as
functions of density with several typical values of $\mu $ and
$g^{2}/\mu ^{2}$. Here, the $E_{\mathrm{sym,UB}}(\rho )$ is
extracted from the parabolic approximation $E_{\mathrm{sym,UB}}(\rho
)\approx E_{\mathrm{UB}}(\rho ,\delta =1)-E_{\mathrm{UB}}(\rho
,\delta =0)$
with $E_{\mathrm{UB}}(\rho ,\delta )$ being the energy per nucleon from the $%
U$-boson contribution. One can see from Fig. \ref{EsymUB} that both $E_{%
\mathrm{0,UB}}^{E}(\rho )$ and $E_{\mathrm{sym,UB}}(\rho )$ increase
with increasing values of both $\mu $ and $g^{2}/\mu ^{2}$. As
expected, however, both the contributions of the exchange term to
energy per nucleon of symmetric nuclear matter and the symmetry
energy are quite small and can be safely neglected compared with the
direct term contribution for the values of $\mu $ and $g^{2}/\mu
^{2}$ considered here. For example, even at very high
baryon density such as $\rho =1.0$ fm$^{-3}$ with $\mu =30.0$ MeV and $%
g^{2}/\mu ^{2}=150$ GeV$^{-2}$ (See the right panels of Fig.
\ref{EsymUB}), the magnitude of the $E_{\mathrm{0,UB}}^{E}(\rho )$
is only about $1.1$ MeV and the magnitude of the
$E_{\mathrm{sym,UB}}(\rho )$ is less than about $0.32$ MeV while the
direct term contribution to the energy per nucleon of symmetric
nuclear matter reaches about $576$ MeV. The magnitudes of
$E_{\mathrm{0,UB}}^{E}(\rho )$ and $E_{\mathrm{sym,UB}}(\rho )$ will
further decrease if smaller values of $\mu $ and $g^{2}/\mu ^{2}$
are used (See, e.g., the left panels of Fig. \ref{EsymUB}). These
results verify the validity of neglecting the exchange term
contribution of the $U$-boson to nuclear matter EOS in the
literature~\cite{Kri09,Wen09,Zha11,Wen11}.

\subsection{The core-crust transition density and pressure in neutron
stars with the $U$-boson}

We now turn to the numerical results on the core-crust transition
density and pressure in neutron stars with the thermodynamical
approach, the curvature-matrix approach, and the Vlasov equation
approach. We note that both the matrices $C_{CM}^{f}$ in
Eq.~(\ref{cur_matrix}) and $C_{VE}^{f}$ in Eq.~(\ref{vla_matrix})
are $k$-dependent, and for the curvature-matrix approach and the
Vlasov equation approach, the core-crust transition density
corresponds to the critical baryon number density above which the
neutron star matter is always stable for all possible values of $k$
while below which one can always find a $k$ value to violate the
stability conditions of the neutron star matter. On the other hand,
for the thermodynamical method, the transition density can be
directly obtained by solving the equation $V_{ther}=0$ (See
Eq.~(\ref{Vther}) for the expression of $V_{ther}$).
\begin{figure}[tbp]
\includegraphics[scale=1.0]{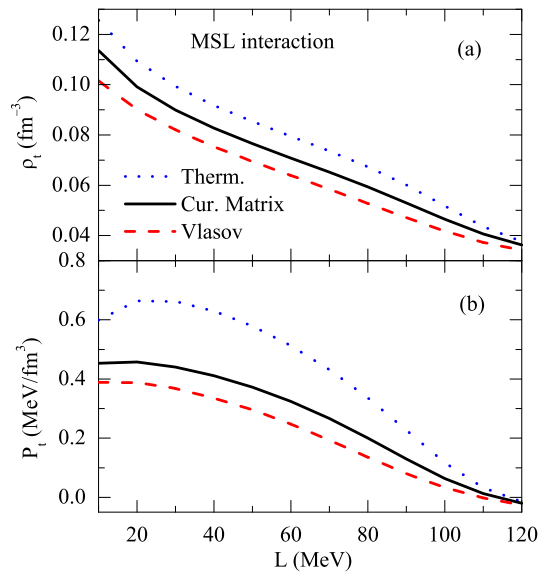}
\caption{(Color online) Transition density $\protect\rho _{t}$ (a) and pressure $%
P_{t}$ (b) in neutron stars as functions of the $L$ parameter with
the MSL interaction using the thermodynamical method, the curvature
matrix method and the Vlasov equation method.} \label{rhotL}
\end{figure}

Theoretically it has been established that there exists a strong correlation
between the transition density $\rho _{t}$ and the nuclear symmetry energy.
In particular, a strong linear correlation between the transition density $%
\rho _{t}$ and the slope parameter $L$ of the nuclear symmetry
energy has been observed in many different theoretical calculations
\cite{XuJ09,Che10,Duc10,Duc11}. To see the symmetry energy
dependence of the inner edge of neutron star crusts, we show in
Fig.~\ref{rhotL} the transition density $\rho _{t}$ and pressure
$P_{t}$ in neutron stars as functions of the $L$ parameter with MSL0
interaction by varying individually $L$ using the thermodynamical
method, the curvature matrix method and the Vlasov equation method.
When varying individually the $L$ parameter, we keep all other
macroscopic quantities $\rho _{0}$, $E_{0}(\rho _{0})$, $K_{0}$,
$m_{s,0}^{\ast }$, $m_{v,0}^{\ast }$, $E_{\text{\textrm{sym}}}({\rho
_{0}})$, $G_{S}$, $G_{V}$, and $W_{0}$ at their default values in
MSL0. It should be noted that the original agreement of MSL0 with
the experimental data of binding energies or charge radii of finite
nuclei essentially still holds with the individual change of the $L$
parameter.

For the results shown in Fig.~\ref{rhotL}, the $U$-boson
contributions are not considered. It is seen that all the three
methods give similar results for the $L$ dependence of the
transition density $\rho _{t}$ and pressure $P_{t}$ with the
curvature-matrix method giving slightly smaller values of $\rho
_{t}$ and $P_{t}$ than the thermodynamical method while slightly
higher values of $\rho _{t}$ and $P_{t}$ than the Vlasov equation
method for a fixed value of $L$. The nonmonotonous variation of the
$L$ dependence of the transition pressure $P_{t}$ in the
thermodynamical method is due to the fact that the $P_{t}$ is a
complicated functions of $\rho _{t}$, $L$, and the isospin asymmetry
$\delta _{t}$ at the transition density (See, e.g., \cite{XuJ09}).
The smaller values of $\rho _{t}$ from the curvature-matrix method
than from the thermodynamical method implies that the density
gradient terms and Coulomb term considered in the former can make
the neutron star matter more stable, which are consistent with the
results in Ref. \cite{XuJ09,Che10} (The curvature-matrix method is
called dynamical method there). On the other hand, the slightly
smaller values predicted by the Vlasov equation method than the
curvature-matrix method is due to the quantum effects considered in
the Vlasov equation method, indicating that the quantum effects will
make the neutron star matter more stable as expected.
\begin{figure}[tbp]
\includegraphics[scale=0.85]{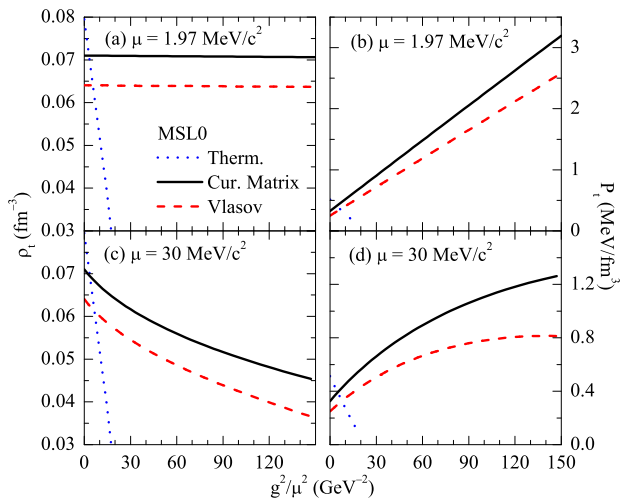}
\caption{(Color online) The $g^{2}/\protect\mu ^{2}$ dependence of
the transition density $\protect\rho _{t}$ and pressure $P_{t}$ in
neutron stars from the thermodynamical method, the curvature matrix
method and the Vlasov equation method with the MSL0 interaction for
$\protect\mu =1.97$ MeV and $30.0$ MeV, respectively.} \label{rhotG}
\end{figure}

To see how the light and weakly coupled vector $U$-boson affects the
inner edge of neutron star crusts, we show in Fig.~\ref{rhotG} the
$g^{2}/\mu ^{2}$ dependence of the transition density $\rho _{t}$
and pressure $P_{t}$ in neutron stars from the MSL0 interaction by
including the $U$-boson with $\mu =1.97$ MeV and $30.0$ MeV,
respectively, using the thermodynamical method, the curvature matrix
method and the Vlasov equation method. One can see clearly that
while the curvature matrix method and the Vlasov equation
method predict very similar results for the $g^{2}/\mu ^{2}$ dependence of $%
\rho _{t}$ and $P_{t}$, the thermodynamical method predicts very different
results with $\rho _{t}$ and $P_{t}$ decreasing very quickly as the $%
g^{2}/\mu ^{2}$ increases. In particular, we find that for
$g^{2}/\mu ^{2}>20 $ GeV$^{-2}$, the neutron star matter is always
stable and the transition density does not exist in the
thermodynamical method. This is due to the fact that the
$k$-dependent terms in Eq.~(\ref{cur_matrix}) and
Eq.~(\ref{vla_matrix}) originating from the finite-range interaction
and density-gradient contributions play an important role in
determining the transition density when the $U$-boson is considered.
Therefore, the thermodynamical method which ignores the
$k$-dependent terms would be no longer appropriate to determine the
inner edge of neutron star crusts when the $U$-boson\ is taken into
account.

It is interesting to see from Fig.~\ref{rhotG} that for the
curvature matrix method and the Vlasov equation method, the effects
of the $U$-boson on the transition density $\rho _{t}$ and pressure
$P_{t}$ depend on not only the ratio $g^{2}/\mu ^{2}$ but also the
$U$-boson mass $\mu $. In particular, for a heavier $U$-boson (e.g.,
$\mu =30.0$ MeV), the $\rho _{t}$ decreases
significantly with increment of $g^{2}/\mu ^{2}$ from $0$ to $150$ GeV$%
^{-2}$. On the other hand, for a very light $U$-boson (e.g., $\mu =1.97$
MeV), the transition density $\rho _{t}$ exhibits very weak dependence on $%
g^{2}/\mu ^{2}$. These features imply that for a fixed value of
$g^{2}/\mu ^{2}$, a heavier $U$-boson can make the neutron star
matter more stable while a very light $U$-boson essentially has no
influence on the transition density.

As shown in Fig.~\ref{rhotG}, for the curvature matrix method and
the Vlasov equation method, although the transition density $\rho
_{t}$ displays a somewhat complicated relationship with the
properties of the $U$-boson, the transition pressure $P_{t}$ simply
increases with the ratio $g^{2}/\mu ^{2}$ whether the $U$-boson mass
is light or heavy. Furthermore, it is interesting to see from
Fig.~\ref{rhotG} (b) that the transition pressure $P_{t}$ increases
almost linearly with $g^{2}/\mu ^{2}$ for a very light $U$-boson
(i.e., $\mu =1.97$ MeV), while it exhibits much slower increment
with $g^{2}/\mu ^{2}$ for a heavier $U$-boson (e.g., $\mu =30.0$
MeV) as shown in Fig.~\ref{rhotG} (d). The linear correlation
between $P_{t}$ and $g^{2}/\mu ^{2}$ for $\mu =1.97$ MeV observed in
Fig.~\ref{rhotG} (b) is easily understood since the $U$-boson
contribution to the pressure $P_{t,\mathrm{UB}}$ is dominated by the
direct term contribution, i.e., $P_{t,\mathrm{UB}}\approx
\frac{1}{2}g^{2}/\mu ^{2}\rho _{t}^{2}$, if\ $\mu $ is independent
of the density as we assume in the present work, and the $\rho _{t}$
remains approximately a constant value for $\mu =1.97$ MeV when
$g^{2}/\mu ^{2}$ varies from $0$ to $150$ GeV$^{-2}$ as shown
Fig.~\ref{rhotG} (a). In the case of $\mu =30.0$ MeV, the $\rho
_{t}$ decreases significantly with the increment of $g^{2}/\mu ^{2}$
as shown Fig.~\ref{rhotG} (c), which leads to the transition
pressure $P_{t}$ displays much slower increment with $g^{2}/\mu
^{2}$ as shown in Fig.~\ref{rhotG} (d).
\begin{figure}[tbp]
\includegraphics[scale=1.0]{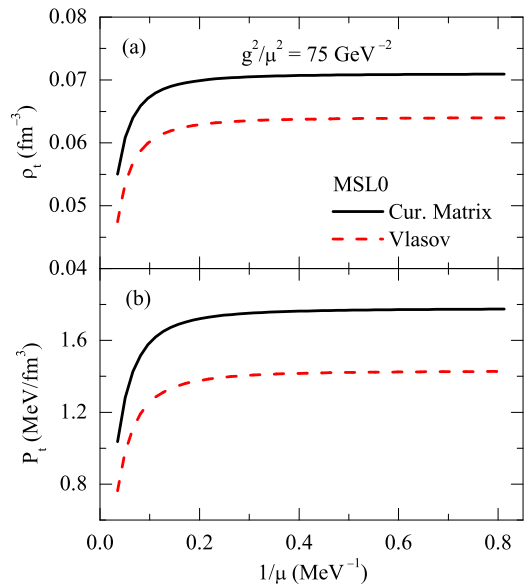}
\caption{(Color online) The $1/\protect\mu $ dependence of the
transition density $\protect\rho _{t}$ and pressure $P_{t}$ in
neutron stars from the the curvature matrix method and the Vlasov
equation method with the MSL0 interaction for $g^{2}/\protect\mu
^{2}=75$ GeV$^{-2}$.} \label{rhotMu}
\end{figure}

In order to see the $U$-boson mass dependence of $\rho _{t}$ and
$P_{t}$ at a fixed value of $g^{2}/\mu ^{2}$, we display in
Fig.~\ref{rhotMu} the $1/\mu $ dependence of the transition density
$\rho _{t}$ and pressure $P_{t}$ in neutron stars from the MSL0
interaction by including the $U$-boson contribution with $g^{2}/\mu
^{2}=75$ GeV$^{-2}$ using the curvature matrix method and the Vlasov
equation method. It is seen that the two methods predict very
similar $1/\mu $ dependence of the transition density $\rho _{t}$
and pressure $P_{t}$ with the Vlasov equation method giving smaller
values. It is interesting to see that $\rho _{t}$ becomes sensitive
to $\mu $ when the $U$-boson mass $\mu $ is larger than about $2$
MeV although the $U$-boson almost has no influence on the transition
density $\rho _{t}$ if its mass $\mu $ is less than about $2$ MeV.
The transition pressure $P_{t}$ displays similar $1/\mu $ dependence
as the $\rho _{t}$ due to the relation $P_{t,\mathrm{UB}}\approx
\frac{1}{2}g^{2}/\mu ^{2}\rho _{t}^{2}$.\ Therefore, these results
demonstrate that the transition density $\rho _{t}$ in neutron stars
can be sensitive to the value of both $\mu $ and $g^{2}/\mu ^{2}$,
and any experimental or observational constraints on $\rho _{t}$ may
put important limits on $\mu $ and $g^{2}/\mu ^{2}$, or equivalently
on $\mu $ and $g$.

\begin{figure}[tbp]
\includegraphics[scale=0.9]{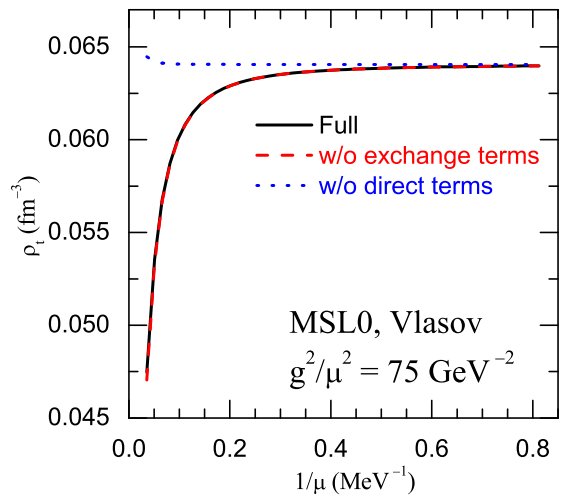}
\caption{(Color online) The $1/\protect\mu $ dependence of the
transition density $\protect\rho _{t}$ in neutron stars from the
Vlasov equation method with the MSL0 interaction for
$g^{2}/\protect\mu ^{2}=75$ GeV$^{-2}$. The results of neglecting
the $U$-boson direct term contribution or neglecting the $U$-boson
exchange term contribution are included for comparison.}
\label{rhotMuEx}
\end{figure}

As have been shown above, the exchange term contribution of the
$U$-boson to the nuclear matter EOS can be safely neglected. It is
thus interesting to see how the exchange term affects the inner edge
of neutron star crusts. To check this point, we show in
Fig.~\ref{rhotMuEx} the $1/\mu $ dependence of the transition
density $\rho _{t}$ in neutron stars with the MSL0 interaction for
$g^{2}/\mu ^{2}=75$ GeV$^{-2}$ from the Vlasov equation method
together with that by neglecting the direct term contribution or
neglecting the exchange term contribution. It is seen that the
exchange term contribution of the $U$-boson has very small influence
on the transition density $\rho _{t}$, especially for the light mass
$U$-boson. On the other hand, the direct term significantly affects
the $1/\mu $ dependence of the transition density $\rho _{t}$.
Particularly, neglecting the direct term contribution of the
$U$-boson leads to very weak $1/\mu $ dependence of the transition
density $\rho _{t}$, implying that the observed strong $1/\mu $
dependence of the transition density $\rho _{t}$ is essentially due
to the direct term contribution which produces a $k$-dependent term
in the matrix (\ref{cur_matrix}) or (\ref{vla_matrix}), i.e., the
third term in the right hand side of Eq.~(\ref{cur_matrix}). We note
that using the curvature matrix method leads to the same conclusion.

\subsection{The mass-radius relation and crustal fraction of moment of
inertia for static neutron stars with the $U$-boson}

As we have showed above, the $U$-boson may have significant
influence on the nuclear matter EOS and the inner edge of neutron
star crusts. Here we investigate effects of the $U$-boson on the
global properties of static neutron stars. To calculate the global
properties, such as the mass-radius relation and crustal fraction of
moment of inertia, of static neutron stars, one needs the EOS of
neutron star matter over a broad density region ranging from the
center to the surface of neutron stars. Besides the possible
appearance of nuclear pasta in the inner crust, various phase
transitions and non-nucleonic degrees of freedom may appear in the
core of neutron stars. In this work, we restrict ourselves to the
simplest and traditional model, and make the minimum assumption that
the core of neutron stars contains the uniform $\beta $-stable and
electrically neutral $npe\mu $ matter only and there is no phase
transition.

Generally, a typical neutron star contains the liquid core, inner
crust and outer crust from the center to surface. For the liquid
core we use the EOS of $npe\mu $ matter from SHF calculations
including the $U$-boson contributions to the nuclear EOS. For the
Skyrme effective nucleon-nucleon interaction, the MSL interaction
with a soft symmetry energy of $L=30$ MeV is used. We note that the
MSL interaction with $L=30$ MeV predicts a $npe\mu $ matter EOS\
very similar to the more sophisticated EOS\ containing nucleons,
hyperons and quark\ degrees of freedom \cite{XuJ10a}. In the present
work, the $U$-boson contribution to the EOS of liquid core includes
both the direct term contribution (i.e., Eq.~(\ref{HUBD})) and the
exchange term contribution (i.e., Eq.~(\ref{HUBE}) without the
density gradient terms) although the latter is negligible. In
particular, the fractions of neutrons, protons, electrons and muons
in the neutron star matter are obtained from self-consistently
solving the set of equations for $\beta $-stable condition (i.e.,
$\mu_{np} =\mu _{e} =\mu _{\mu}$) and charge neutral condition(i.e.,
$\rho_{p} =\rho _{e} + \rho _{\mu}$) by considering the $U$-boson
exchange term contribution to the chemical potential of neutrons and
protons. It should be noted that the $U$-boson direct term does not
change the neutron and proton chemical potential difference
$\mu_{np}$ as it contributes equally to the single-particle
potential of neutrons and protons, and thus the chemical
compositions of the neutron star matter will not change if only the
$U$-boson direct term contribution is considered as pointed out in
previous work~\cite{Kri09,Wen09,Zha11,Wen11}. In this way, the
contributions of $U$-boson to the energy density, the pressure, the
nucleon effective masses and chemical potentials, are considered
self-consistently in the neutron star matter calculations for the
liquid core.

\begin{figure}[tbp]
\includegraphics[scale=0.85]{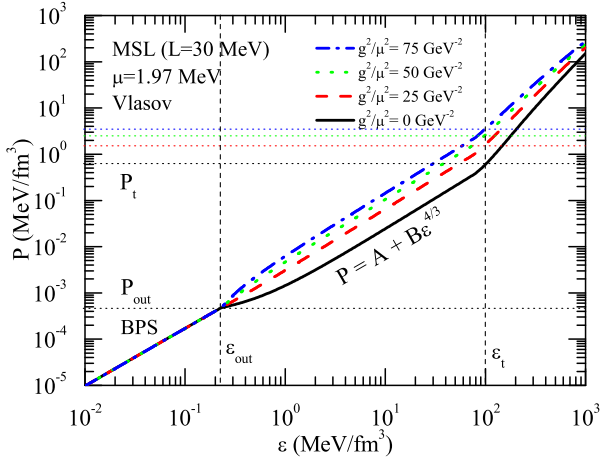}
\caption{(Color online) The EOS's of different parts in neutron
stars using the MSL interaction with $L=30$ MeV for different values
of $g^{2}/\protect\mu ^{2}$ ranging from $0$ to $75$ GeV$^{-2}$ with
$\mu = 1.97$ MeV. The energy density (pressure) at $\rho_t$ and
$\rho_{out}$ is indicated as $\varepsilon_t$ ($P_{t}$) and
$\varepsilon_{out}$ ($P_{out}$), respectively, and the $\rho _{t}$
is obtained from the Vlasov equation method.} \label{Pepsilon}
\end{figure}

In the inner crust with densities between $\rho _{out}$ and $\rho
_{t}$ where the nuclear pastas may exist, because of our poor
knowledge about its EOS from first principle, following Carriere
\textsl{et al.} \cite{Hor03} (See also Ref.~\cite{XuJ09}) we
construct its EOS according to
\begin{equation}
P=a+b\epsilon ^{4/3}.  \label{crustEOS43}
\end{equation}%
This polytropic form with an index of $4/3$ has been found to be a good
approximation to the crust EOS \cite{Lin99,Lat00}. The $\rho
_{out}=2.46\times 10^{-4}$ fm$^{-3}$ is the density separating the inner
from the outer crust. The constants $a$ and $b$ are then determined by
\begin{eqnarray}
a &=&\frac{P_{out}\epsilon _{t}^{4/3}-P_{t}\epsilon _{out}^{4/3}}{\epsilon
_{t}^{4/3}-\epsilon _{out}^{4/3}}, \\
b &=&\frac{P_{t}-P_{out}}{\epsilon _{t}^{4/3}-\epsilon _{out}^{4/3}},
\end{eqnarray}%
where $P_{t}$, $\epsilon _{t}$ and $P_{out}$, $\epsilon _{out}$ are
the pressure and energy density at $\rho _{t}$ and $\rho _{out}$,
respectively. In the outer crust with $6.93\times 10^{-13}$
fm$^{-3}<\rho <\rho _{out}$, we use the EOS of BPS
\cite{BPS71,Iida97}, and in the region of $4.73\times 10^{-15}$
fm$^{-3}<\rho <$$6.93\times 10^{-13}$ fm$^{-3}$ we use the EOS of
Feynman-Metropolis-Teller (FMT)~\cite{BPS71}. For the $U$-boson
contribution to the EOS of neutron star crusts and surface, we add
the energy density and pressure from the $U$-boson direct term
contribution, i.e., $\epsilon _{\mathrm{UB}}=P_{\mathrm{UB}}\approx
\frac{1}{2}g^{2}/\mu ^{2}\rho ^{2}$ to the corresponding parts since
the exchange term contribution is negligible as shown in the above.

As an example, we show in Fig.~\ref{Pepsilon} the EOS for different
parts of a neutron star. As we have discussed earlier, the
transition density $\rho _{t} $ is obtained by studying the onset of
instabilities in the liquid core, namely it is the critical density
below which small density fluctuations will grow exponentially.
Therefore, the transition density $\rho _{t}$ and the EOS of the
liquid core are obtained self-consistently from the same interaction
and in this sense they are on the same footing. We use in
Fig.~\ref{Pepsilon} the $\rho _{t}$ obtained within the Vlasov
equation method using the full EOS with the MSL interaction of
$L=30$ MeV for different values of $g^{2}/\mu ^{2}$ with $\mu =
1.97$ MeV. Using the above EOS for different parts of the neutron
star, the radial distribution of the total energy density and the
pressure in neutron stars is continuous, but the derivative of the
pressure is not continuous at $\rho _{t}$ and $\rho _{out}$. It is
seen that the EOS's of the inner crust and the liquid core are quite
different for different values of $g^{2}/\mu ^{2}$. Interestingly,
one can see that because the transition density $\rho _{t}$ displays
very weak dependence on $g^{2}/\mu ^{2}$ for the very light
$U$-boson mass ($1.97$ MeV here), the $\rho _{t}$ (and thus the
corresponding $\varepsilon_t$) has essentially the same value for
different $g^{2}/\mu ^{2}$, and the difference of the EOS for
different values of $g^{2}/\mu ^{2}$ observed in Fig.~\ref{Pepsilon}
is essentially due to the variation of $P_{t}$ with $g^{2}/\mu ^{2}$
because of the relation $P_{t,\mathrm{UB}}\approx
\frac{1}{2}g^{2}/\mu ^{2}\rho _{t}^{2}$.

\begin{figure}[tbp]
\includegraphics[scale=1.0]{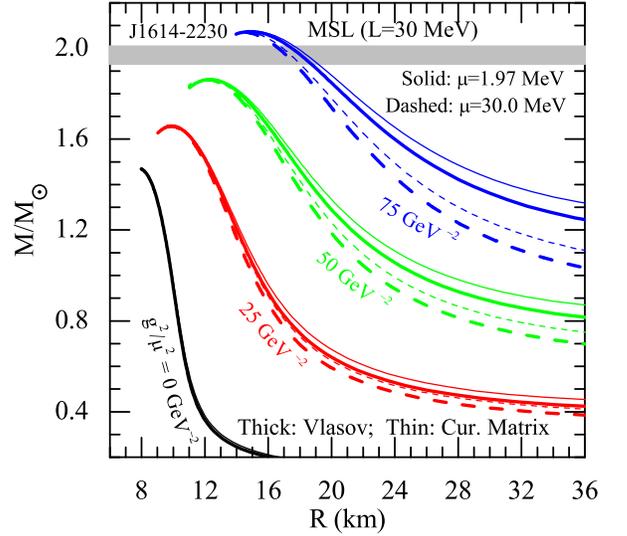}
\caption{(Color online) The mass-radius relation of static neutron
stars using the MSL interaction with $L=30$ MeV and different values
of $g^{2}/\protect\mu ^{2}$ for $\protect\mu =1.97$ MeV (solid
lines) and $30.0$ MeV (dashed lines), respectively. The results with
$\rho _{t}$ from both the Vlasov equation method (thick lines) and
the curvature matrix method (thin lines) are included for
comparison. The shaded band represents the latest new holder of the
maximum mass of neutron stars of $1.97\pm 0.04M_{\odot }$ from PSR
J1614-2230 \cite{Dem10}.} \label{MR}
\end{figure}

Using the EOS constructed above, one can solve the
Tolman-Oppenheimer-Volkoff (TOV) equations to obtain the mass-radius
relations and the results are shown in Fig. \ref{MR}. Indicated by
the shaded band in Fig. \ref{MR} is the latest new holder of the
maximum mass of neutron stars of $1.97\pm 0.04M_{\odot }$ from PSR
J1614-2230 \cite{Dem10}. For MSL interaction with a soft symmetry
energy ($L=30$ MeV) without considering the $U$-boson contribution,
the neutron star mass $M$ decreases quickly with increasing radius
$R$ and the maximum mass is about $1.47$ $M_{\odot }$, which is
significantly less than the observed maximum neutron star mass of
$1.97\pm 0.04M_{\odot }$. On the other hand, the neutron star mass
can be enhanced strongly if the effects of $U$-boson are considered.
In particular, a larger value of $g^{2}/\mu ^{2}$ leads to a larger
neutron star mass at a fixed radius since the nuclear EOS is
increasingly stiffened with increment of $g^{2}/\mu ^{2}$ as shown
in Fig.~\ref{Pepsilon}. Especially, the neutron star maximum mass
can reach $2.07M_{\odot }$ with $g^{2}/\mu ^{2}=75$ GeV$^{-2}$, and
the corresponding radius of the maximum mass neutron star is about
$14.8$ km.

To see more clearly the $U$-boson effects on the properties of
neutron stars, we include in Fig.~\ref{MR} the mass-radius relations
for different values of $g^{2}/\mu ^{2}$ with two values of the
$U$-boson mass, i.e., $\mu =1.97$ and $30.0$ MeV, respectively.
Furthermore, the results with $\rho _{t}$ from both the Vlasov
equation method and the curvature matrix method are included for
comparison although the former is believed to be more realistic. As
expected, for the larger value of the $U$-boson mass, a smaller
value of the neutron star mass $M$ at a fixed radius $R$ is obtained
due to the smaller transition density and pressure obtained for a
larger $\mu $ as shown in the previous section. This $U$-boson mass
effect on the neuron star mass-radius relation will become more
pronounced for a larger value of $g^{2}/\mu ^{2}$ (e.g., $g^{2}/\mu
^{2}=75$ GeV$^{-2}$). In addition, one can see that using the $\rho
_{t}$ from the curvature matrix method predicts a little larger
neutron star mass $M$ at a fixed radius $R$ than using the $\rho
_{t}$ from the Vlasov equation method. However, the maximum mass of
the neutron stars is almost the same for different values of $\mu $
using the $\rho _{t}$ from either the curvature matrix method or the
Vlasov equation method. These results indicate that, essentially
regardless of the values of the $U$-boson mass and the $\rho _{t}$,
a value of $g^{2}/\mu ^{2}=75$ GeV$^{-2}$ can reasonably describe
the latest new holder of the neutron star maximum mass of $1.97\pm
0.04M_{\odot }$ from PSR J1614-2230 \cite{Dem10}.

\begin{figure}[tbp]
\includegraphics[scale=0.95]{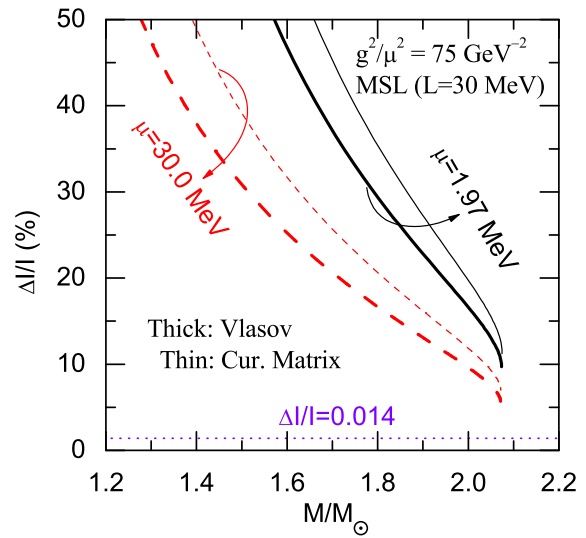}
\caption{(Color online) The crustal fraction of total moment of
inertia $\protect\Delta I/I$ as a function of the neutron star mass
for static neutron stars using the MSL interaction with $L=30$ MeV
for $g^{2}/\protect\mu ^{2} = 75$ GeV$^{-2}$ with $\mu = 1.97$ MeV
(solid lines) and $30.0$ MeV (dashed lines), respectively. The
results with $\rho _{t}$ from both the Vlasov equation method (thick
lines) and the curvature matrix method (thin lines) are included for
comparison. The lower limit $\protect\Delta I/I =0.014$ of the
observation constraint for the Vela pulsar~\cite{Lin99} is also
indicated.} \label{dIM}
\end{figure}

The crustal fraction of total moment of inertia of a static neutron
star, $\Delta I/I$, is a particularly interesting quantity as it can
be inferred from observations of pulsar glitches, i.e., the
occasional disruptions of the otherwise extremely regular pulsations
from magnetized, rotating neutron stars. Furthermore, as was
stressed by Lattimer and Prakash~\cite{Lat00}, the $\Delta I/I$
depends sensitively on the $\rho _{t}$ and $P_{t}$ which are
essentially determined by the EOS of asymmetric nuclear matter at
subsaturation densities as shown above, but there is no explicit
dependence upon the EOS of neutron star matter at higher densities.
These features imply that the $\Delta I/I$ may provide a good probe
for properties of the $U$-boson. To illustrate the $U$-boson mass
dependence of the $\Delta I/I$, we show in Fig. \ref{dIM} the
$\Delta I/I$ as a function of the neutron star mass for static
neutron stars using the MSL interaction with $L=30$ MeV for
$g^{2}/\mu ^{2} = 75$ GeV$^{-2}$ with $\mu = 1.97$ and $30.0$ MeV,
respectively. Furthermore, the results with $\rho _{t}$ obtained
from both the Vlasov equation method and the curvature matrix method
are included for comparison. The $\Delta I/I$ is obtained here from
direct numerical calculations as in Ref.~\cite{XuJ09}. As expected,
one can see that the $\Delta I/I$ indeed exhibits a clear
sensitivity to the $U$-boson mass with a heavier $U$-boson mass
giving a smaller $\Delta I/I$ for a fixed neutron star mass. In
addition, one can see that using the $\rho _{t}$ obtained from the
curvature matrix method gives a little larger value for $\Delta I/I$
at a fixed neutron star mass $M$ than using the $\rho _{t}$ from the
Vlasov equation method. Empirically, the crustal fraction of total
moment of inertia has been constrained as $\Delta I/I > 0.014$ from
studying the glitches of the Vela pulsar~\cite{Lin99} and the lower
limit $\protect\Delta I/I =0.014$ of the constraint is also
indicated in Fig. \ref{dIM}. It is seen from Fig. \ref{dIM} that the
calculated results of $\Delta I/I$ with the two values of $\mu =
1.97$ and $30.0$ MeV using the $\rho _{t}$ obtained from either the
curvature matrix method or the Vlasov equation method are all
consistent with the observation constraint of $\Delta I/I > 0.014$
for the Vela pulsar.

Based on the results above, we conclude that the vector $U$-boson
can significantly stiffen the nuclear matter EOS and thus enhance
strongly the (maximum) mass of neutron stars. Furthermore, as the
$\rho _{t}$ and $P_{t}$ are sensitive to both $g^{2}/\mu ^{2}$ and
$\mu $ if the $U$-boson mass is larger than about $2$ MeV, both
$g^{2}/\mu ^{2}$ and $\mu $ can thus affect the mass-radius relation
of neutron stars although the maximum mass of neutron stars is
essentially independent of the $U$-boson mass $\mu $. In addition,
our results demonstrate that the crustal fraction of total moment of
inertia $\Delta I/I$ may depend sensitively on the $U$-boson mass
$\mu $ for a fixed value of $g^{2}/\mu ^{2}$.

\section{Summary}

Using the thermodynamical approach, the curvature matrix approach
and the Vlasov equation approach with the Skyrme effective
nucleon-nucleon interaction, we have investigated effects of the
light vector gauge $U$-boson, that is weakly coupled to nucleons, on
the core-crust transition density $\rho _{t}$ and pressure $P_{t}$
of neutron stars. For the exchange term contribution of the
$U$-boson, we have applied the density matrix expansion approach,
which automatically leads to the density gradient terms in the
single nucleon potential and nuclear energy density functional. Our
results have demonstrated that the exchange term contribution to
energy per nucleon of symmetric nuclear matter and the symmetry
energy is quite small and can be safely neglected compared with the
direct term contribution for the parameter range of $\mu $ and
$g^{2}/\mu ^{2}$ considered in this work, verifying the validity of
neglecting the $U$-boson exchange term contribution to the nuclear
matter EOS as have been done in the literature. Furthermore, the
exchange term has also been found to have negligible influence on
the core-crust transition density $\rho _{t}$ and pressure $P_{t}$,
especially for very light $U$-boson.

Interestingly, our results have shown that the $\rho _{t}$ and
$P_{t}$ depend on not only the ratio of coupling strength to mass
squared of the $U$-boson $g^{2}/\mu ^{2}$ but also its mass $\mu $.
The $U$-boson mass dependence of $\rho _{t}$ and $P_{t}$ is due to
the finite range interaction from the $U$-boson exchange.
Especially, we have found that the $\rho _{t}$ and $P_{t}$ will be
sensitive to both $g^{2}/\mu ^{2}$ and $\mu $ if the $U$-boson mass
$\mu $ is larger than about $2$ MeV, and both $g^{2}/\mu ^{2}$ and
$\mu $ can have significant influence on the mass-radius relation
and the crustal fraction of total moment of inertia of neutron
stars. Therefore, our results presented in this work have
demonstrated that astrophysical observations on neutron star
structures, such as the mass-radius relation and the crustal
fraction of total moment of inertia from pulsar glitches, can be
potentially useful to constrain properties of the $U$-boson, e.g.,
its mass $\mu $ and the coupling constant $g$ to nucleons.

\section*{ACKNOWLEDGMENTS}

The authors would like to thank W.Z. Jiang, B.A. Li, D.H. Wen, and
J. Xu for useful discussions. This work was supported in part by the
National Natural Science Foundation of China under Grant Nos.
10975097 and 11135011, Shanghai Rising-Star Program under grant No.
11QH1401100, ``Shu Guang" project supported by Shanghai Municipal
Education Commission and Shanghai Education Development Foundation,
and the National Basic Research Program of China (973 Program) under
Contract No. 2007CB815004.


\begin{thebibliography}{99}
\bibitem{Fay80} P. Fayet, Phys. Lett. 95\textbf{B}, 285 (1980); Nucl. Phys.
B \textbf{187}, 184 (1981).

\bibitem{Jean03} P. Jean \textit{et al.}, A\&A \textbf{407}, L55 (2003).

\bibitem{Boe04a} C. Boehm, D. Hooper, J. Silk, M. Casse, and J. Paul, Phys.
Rev. Lett. \textbf{92}, 101301 (2004).

\bibitem{Boe04b} C. Boehm, P. Fayet, and J. Silk, Phys. Rev. D \textbf{69},
101302(R) (2004).

\bibitem{Boe04c} C. Boehm and P. Fayet, Nucl. Phys. B \textbf{683},
291 (2004).

\bibitem{Fuj71} Y. Fujii, Nature (London), Phys. Sci. \textbf{234}, 5 (1971).

\bibitem{Ark98} N. Arkani-Hamed et al., Phys Lett. \textbf{B429}, 263
(1998); Phys. Rev. \textbf{D59}, 086004 (1999).

\bibitem{Fis99} E. Fischbach and C.L. Talmadge, The Search for Non-Newtonian
Gravity, Springer-Verlag, New York, Inc. (1999), ISBN 0-387-98490-9.

\bibitem{Pea01} R. Pease, Nature \textbf{411}, 986 (2001).

\bibitem{Hoy03} C.D. Hoyle, Nature \textbf{421}, 899 (2003).

\bibitem{Lon03} J.C. Long \textit{et al.}, Nature \textbf{\ 421}, 922 (2003).

\bibitem{Ade03} E. G. Adelberger \textit{et al.}, Annu. Rev. Nucl. Part.
Sci. \textbf{53}, 77 (2003); Prog. Part. Nucl. Phys. \textbf{62},
102 (2009).

\bibitem{Uza03} J.P. Uzan, Rev. Mod. Phys. \textbf{75}, 403 (2003).

\bibitem{Dec05} R.S. Decca et al., Phys. Rev. Lett. \textbf{94}, 240401
(2005).

\bibitem{Rey05} S. Reynaud \textit{et al.}, Int. J. Mod. Phys. \textbf{A20},
2294 (2005).

\bibitem{Pok06} Y.N. Pokotilovski, Phys. At. Nucl. \textbf{69}, 924 (2006).

\bibitem{Kap07} D.J. Kapner \textit{et al.}, Phys. Rev. Lett. \textbf{98}, 021101
(2007).

\bibitem{Nes08} V.V. Nesvizhevsky \textit{et al.}, Phys. Rev. D \textbf{77}%
, 034020 (2008).

\bibitem{Kam08} Y. Kamyshkov, J. Tithof, and M. Vysotsky, Phys. Rev. D
\textbf{78}, 114029 (2008).

\bibitem{Aza08} M. Azam, M. Sami, C.S. Unnikrishnan, and T. Shiromizu, Phys.
Rev. D \textbf{77}, 101101 (2008).

\bibitem{New09} R.D. Newman, E.C. Berg, and P.E. Boynton, Space Sci.
Rev. \textbf{148}, 175 (2009).

\bibitem{Ger10} A. A. Geraci \textit{et al.}, Phys. Rev. Lett. \textbf{105},
101101 (2010).

\bibitem{Luc10} D. M. Lucchesi and R. Person, Phys. Rev. Lett. \textbf{105},
231103 (2010).

\bibitem{Fay07} P. Fayet, Phys. Rev. D \textbf{75}, 115017 (2007).

\bibitem{Zhu07} S.H. Zhou, Phys. Rev. D \textbf{75}, 115004 (2007).

\bibitem{Che08} C.H. Chen, C.Q. Geng, and C.W. Kao, Phys. Lett. \textbf{B663},
400 (2008).

\bibitem{Fay09} P. Fayet, Phys. Lett. \textbf{B675}, 267 (2009).

\bibitem{Bar75} R. Barbieri and T.E.O. Ericson, Phys. Lett. \textbf{57B},
270 (1975).

\bibitem{Nes04} V.V. Nesvizhevsky and K.V. Protasov, Class. Quan. Grav.
\textbf{21}, 4557 (2004).

\bibitem{Kri09} M.I. Krivoruchenko, F. \v{S}imkovic, and A. Faessler, Phys. Rev. D \textbf{79},
125023 (2009).

\bibitem{Wen09} D.H. Wen, B.A. Li, and L.W. Chen, Phys. Rev. Lett. \textbf{103%
}, 211102 (2009).

\bibitem{Zha11} D.R. Zhang, P.L. Yin, W. Wang, Q.C. Wang, and W.Z. Jiang,
Phys. Rev. C \textbf{83}, 035801 (2011).

\bibitem{Wen11} D.H. Wen, B.A. Li, and L.W. Chen, arXiv:1101.1504.

\bibitem{Rei07} W. Reisdorf \textsl{et al}. (FOPI Collaboration), Nucl.
Phys. \textbf{A781}, 459 (2007).

\bibitem{Xia09} Z.G. Xiao, B.A. Li, L.W. Chen, G.C. Yong, and M. Zhang, Phys. Rev. Lett. \textbf{102},
062502 (2009).

\bibitem{Dem10} P.B. Demorest, T. Pennucci, S.M. Ransom, M.S.E.
Roberts, and J.W.T. Hessels, Nature \textbf{467}, 1081 (2010).

\bibitem{XuJ10a} J. Xu, L.W. Chen, C.M. Ko, and B.A. Li, Phys. Rev. C
\textbf{81}, 055803 (2010).

\bibitem{BPS71} G. Baym, C. Pethick, and P. Sutherland, Astrophys. J. \textbf{%
170}, 299 (1971).

\bibitem{BBP71} G. Baym, H.A. Bethe, and C.J. Pethick, Nucl. Phys. \textbf{%
A175}, 225 (1971).

\bibitem{Pet95a} C.J. Pethick and D.G. Ravenhall, Ann. Rev. Nucl. Part.
Sci. \textbf{45}, 429 (1995).

\bibitem{Pet95b} C.J. Pethick, D.G. Ravenhall, and C.P. Lorenz, Nucl.
Phys. \textbf{A584}, 675 (1995).

\bibitem{Lat00} J.M. Lattimer and M. Prakash, Phys. Rep. \textbf{333-334},
121 (2000); Astrophys. J. \textbf{550}, 426 (2001).

\bibitem{Lat07} J.M. Lattimer and M. Prakash, Phys. Rep. \textbf{442}, 109
(2007).

\bibitem{Ste05} A.W. Steiner, M. Prakash, J.M. Lattimer, and P.J. Ellis,
Phys. Rep. \textbf{410}, 325 (2005).

\bibitem{Lin99} B. Link, R.I. Epstein, and J.M. Lattimer, Phys. Rev. Lett.
\textbf{83}, 3362 (1999).

\bibitem{Hor04} C.J. Horowitz \textit{et al.}, Phys. Rev. C \textbf{69}, 045804
(2004); C.J. Horowitz \textit{et al.}, Phys. Rev. C \textbf{70},
065806 (2004)

\bibitem{Bur06} A. Burrows, S. Reddy, and T.A. Thompson, Nucl. Phys.
\textbf{A777}, 356 (2006).

\bibitem{Owe05} B.J. Owen, Phys. Rev. Lett. \textbf{95}, 211101 (2005).

\bibitem{Rus06} S.B. Ruster, M. Hempel, and J. Schaffner-Bielich, Phys.
Rev. C \textbf{73}, 035804 (2006).

\bibitem{Neg72} J.W. Negele and D. Vautherin, Phys. Rev. C \textbf{5}, 1472
(1972); \textbf{11}, 1031 (1975).

\bibitem{XuJ10b} J. Xu and C.M. Ko, Phys. Rev. C \textbf{82}, 044311 (2010).

\bibitem{LCK08} B.A. Li, L.W. Chen, and C.M. Ko, Phys. Rep. \textbf{464},
113 (2008).

\bibitem{You99} D.H. Youngblood, H.L. Clark, and Y.-W. Lui, Phys. Rev. Lett.
\textbf{82}, 691 (1999).

\bibitem{Lui04} Y.-W. Lui, D.H. Youngblood, Y. Tokimoto, H.L. Clark, and B.
John, Phys. Rev. C \textbf{70}, 014307 (2004).

\bibitem{Ma02} Z.Y. Ma \textit{et al.}, Nucl. Phys. \textbf{A703}, 222 (2002).

\bibitem{Vre03} D. Vretenar, T. Niksic, and P. Ring, Phys. Rev. C \textbf{68}%
, 024310 (2003).

\bibitem{Col04} G. Colo, N. Van Giai, J. Meyer, K. Bennaceur, and P. Bonche,
Phys. Rev. C \textbf{70}, 024307 (2004).

\bibitem{Shl06} S. Shlomo, V.M. Kolomietz, and G Col\`{o}, Eur. Phys. J. A
\textbf{30}, 23 (2006).

\bibitem{LiT07} T. Li \textit{et al.}, Phys. Rev. Lett. \textbf{99}, 162503 (2007).

\bibitem{Gar07} U. Garg {et al.}, Nucl. Phys. \textbf{A788}, 36 (2007).

\bibitem{Col09} G. Colo, 2009, arXiv:0902.3739v1 [nucl-th].

\bibitem{Che11b} L.W. Chen and J.Z. Gu, J. Phys. G \textbf{39},
035104 (2012) [arXiv:1104.5407].

\bibitem{Dan02a} P. Danielewicz, R. Lacey, and W.G. Lynch, Science \textbf{%
298}, 1592 (2002).

\bibitem{Aic85} J. Aichelin and C.M. Ko, Phys. Rev. Lett. \textbf{55}, 2661
(1985).

\bibitem{Fuc06a} C. Fuchs, Prog. Part. Nucl. Phys. \textbf{56}, 1 (2006).

\bibitem{Mey66} W.D. Myers and W.J. Swiatecki, Nucl. Phys. \textbf{81}, 1
(1966).

\bibitem{Pom03} K. Pomorski and J. Dudek, Phys. Rev. C \textbf{67}, 044316
(2003).

\bibitem{Bar05} V. Baran, M. Colonna, V. Greco, and M. Di Toro, Phys. Rep.
\textbf{410}, 335 (2005).

\bibitem{Che05} L.W. Chen, C.M. Ko, and B.A. Li, Phys. Rev. Lett.
\textbf{94}, 032701 (2005); Phys. Rev. C \textbf{72}, 064309 (2005);
B.A. Li and L.W. Chen, Phys. Rev. C \textbf{72}, 064611 (2005).

\bibitem{Tsa09} M.B. Tsang, Y. Zhang, P. Danielewicz, M. Famiano, Z.
Li, W. G. Lynch, and A. W. Steiner, Phys. Rev. Lett. \textbf{102},
122701 (2009).

\bibitem{Cen09} M. Centelles, X. Roca-Maza, X. Vi\~{n}as, and M.
Warda, Phys. Rev. Lett \textbf{102}, 122502 (2009); M. Warda, X.
Vi\~{n}as, X. Roca-Maza, and M. Centelles, Phys. Rev. C \textbf{80},
024316 (2009).

\bibitem{XuC10} C. Xu, B.A. Li, and L.W. Chen, Phys. Rev. C \textbf{82},
054607 (2010).

\bibitem{Che11a} L.W. Chen, Phys. Rev. C \textbf{83}, 044308 (2011).

\bibitem{Fen10} Z.Q. Feng and G.M. Jin, Phys. Lett. \textbf{B683},
140 (2010).

\bibitem{Rus11} P. Russotto \textit{et al.}, Phys. Lett. \textbf{B697},
471 (2011).

\bibitem{Cha97} E. Chabanat, P. Bonche, P. Haensel, J. Meyer, and R.
Schaeffer, Nucl. Phys. \textbf{A627}, 710 (1997).

\bibitem{Bra85} M. Brack, C. Guet, and H.-B. H\aa kansson, Phys. Rep.
\textbf{123}, 275 (1985).

\bibitem{Rav83} D.G. Ravenhall, C.J. Pethick, and J.R. Wilson, Phys. Rev.
Lett. \textbf{50}, 2066 (1983).

\bibitem{Oya93} K. Oyamatsu, Nucl. Phys. \textbf{A561}, 431 (1993).

\bibitem{Lat04} J.M. Lattimer and M. Prakash, Science \textbf{304}, 536
(2004).

\bibitem{Ste08} A.W. Steiner, Phys. Rev. C \textbf{77}, 035805 (2008).

\bibitem{Dou00} F. Douchin and P. Haensel, Phys. Lett. \textbf{B485}, 107
(2000).

\bibitem{Dou01} F. Douchin and P. Haensel, A\&A \textbf{380}, 151 (2001).

\bibitem{Hor03} J. Carriere, C.J. Horowitz, and J. Piekarewicz, Astrophys.
J. \textbf{593}, 463 (2003).

\bibitem{Kub07} S. Kubis, Phys. Rev. C \textbf{76}, 035801 (2007); Phys.
Rev. C \textbf{70}, 065804 (2004).

\bibitem{Wor08} A. Worley, P.G. Krastev, and B.A. Li, Astrophys. J. \textbf{%
685}, 390 (2008).

\bibitem{Oya07} K. Oyamatsu and K. Iida, Phys. Rev. C \textbf{75}, 015801
(2007).

\bibitem{Duc07} C. Ducoin, Ph. Chomaz, and F. Gulminelli, Nucl. Phys. \textbf{%
A789}, 403 (2007).

\bibitem{Cho04} P. Chomaz, M. Colonna, and J. Randrup, Phys. Rep. \textbf{389%
}, 263 (2004).

\bibitem{Pro06} C. Provid\^{e}ncia, L. Brito, S.S. Avancini, D.P. Menezes,
and P. Chomaz, Phys. Rev. C \textbf{73}, 025805 (2006).

\bibitem{Duc08a} C. Ducoin, J. Margueron, and P. Chomaz, Nucl. Phys. \textbf{%
A809}, 30 (2008).

\bibitem{Duc08b} C. Ducoin, C. Provid\^{e}ncia, A.M. Santos, L. Brito, and
P. Chomaz, Phys. Rev. C \textbf{78}, 055801 (2008).

\bibitem{Pai10} H. Pais, A. Santos, L. Brito, and C. Provid\^{e}ncia, Phys.
Rev. C \textbf{82}, 025801 (2010).

\bibitem{Hor01} C.J. Horowitz and J. Piekarewicz, Phys. Rev. Lett. \textbf{86%
}, 5647 (2001); Phys. Rev. C \textbf{64}, 062802(R) (2001); Phys.
Rev. C \textbf{66}, 055803 (2002).

\bibitem{Cal85} H.B. Callen, Thermodynamics, Wiley, New York, 1985.

\bibitem{XuJ09} J. Xu, L. W. Chen, B.A. Li, and H.R. Ma, Phys. Rev. C \textbf{%
79}, 035802 (2009); Astrophys. J. \textbf{697}, 1549 (2009).

\bibitem{Che10} L.W. Chen, C.M. Ko, B.A. Li, and J. Xu, Phys. Rev. C \textbf{82},
024321 (2010).

\bibitem{Duc10} C. Ducoin, J. Margueron, and C. Provid\^{e}ncia,
EPL \textbf{91}, 32001 (2010).

\bibitem{Duc11} C. Ducoin, J. Margueron, C. Provid\^{e}ncia, and I. Vida\~{n}a,
Phys. Rev. C \textbf{83}, 045810 (2011).

\bibitem{Che11} L.W. Chen, Sci. China: Phys. Mech. Astron. \textbf{54},
suppl. 1, s124 (2011) [arXiv:1101.2384].

\bibitem{Jac07} C. Jacoby and S. Nussinov, J. High Energy Phys. \textbf{5},
017 (2007).

\bibitem{Fuj88} Y. Fujii, in Large Scale Structures of the Universe, page
471-477, Eds. J. Audouze et al. (1988), International Astronomical Union.

\bibitem{Iida97} K. Iida and K. Sato, Astrophys. J. \textbf{477}, 294 (1997).
\end{thebibliography}
\end{document}